\documentclass[cmp,final]{svjour}

\usepackage{amsmath,amsfonts,amssymb}
\spnewtheorem*{theorem*}{Theorem}{\bf}{\it}
\spnewtheorem*{lemma*}{Lemma}{\bf}{\it}
\author{ A. Elgart \inst{1} \and  G. M. Graf \inst{2} \and J. H. Schenker \inst{2}}
\institute{Department of Mathematics, Stanford University, Stanford, CA
94305-2125 \and Theoretische Physik, ETH Z\"urich, CH-8093 Z\"urich,
Switzerland }
\title{Equality of the bulk and edge Hall conductances in a mobility gap}%
\date{September 8, 2004}
\numberwithin{equation}{section}
\newcommand{\eq}[1]{eq.~(\ref{#1})}  
\newcommand{\Ev}[1]{\E \left( #1 \right)}  
\newcommand{\norm}[1]{\left\Vert#1\right\Vert}
\newcommand{\abs}[1]{\left\vert#1\right\vert}
\newcommand{\set}[1]{\left\{#1\right\}}
\newcommand{\com}[2]{\left[ #1 , #2 \right ]}
\newcommand{\bra}[1]{\left < #1 \right |}
\newcommand{\ket}[1]{\left | #1 \right >}

\def\eps{\varepsilon}

\def\e{\mathrm e}
\def\im{\mathrm i}
\def\Im{\mathrm{Im}}
\def\half {\frac{1}{2}}
\def\1{{\mathsf 1}}
\def\di{\mathrm d}

\def\Z{\mathbb Z}
\def\N{\mathbb N}
\def\R{\mathbb R}
\def\C{\mathbb C}
\def\E{\mathbb E}
\def\J{\mathcal J}
\def\I{\mathfrak I}

\def\ran{\operatorname{ran}}
\def\Ind{\operatorname{Ind}}
\def\dist{\operatorname{dist}}   
\def\dim{\operatorname{dim}}  
\def\tr{\operatorname{tr}}    
\def\supp{\operatorname{supp}}

\def\Area{\operatorname{Area}}

\def\Re{\operatorname{Re}}
\def\Im{\operatorname{Im}}

\begin{document}

\maketitle
\begin{abstract}
We consider the edge and bulk conductances for 2D quantum Hall systems in which
the Fermi energy falls in a band where bulk states are localized. We show that
the resulting quantities are equal, when appropriately defined. An appropriate
definition of the edge conductance may be obtained through a suitable time
averaging procedure or by including a contribution from states in the localized
band. In a further result on the Harper Hamiltonian, we show that this
contribution is essential. In an appendix we establish quantized plateaus for
the conductance of systems which need not be translation ergodic.
\end{abstract}
\section{Introduction}

Two conductances, $\sigma_B$ and $\sigma_E$, are associated to the Quantum Hall
Effect (QHE), depending on whether the currents are ascribed to the bulk or to
the edge. The equality $\sigma_B = \sigma_E$, suggested by Halperin's analysis
\cite{Ha82} of the Laughlin argument \cite{La81}, has been established in the
context of an effective field theory description \cite{FS93}. It was later
derived in a microscopic treatment of the integral QHE \cite{SBKR00,EG02,Ma03}
for the case that the Fermi energy lies in a spectral gap $\Delta$ of the
single-particle Hamiltonian $H_B$. We prove this equality, by quite different
means, in the more general setting that $H_B$ exhibits Anderson localization in
$\Delta$ \textemdash more precisely, dynamical localization (see \eqref{eq:2}
below). The result applies to Schr\"odinger operators which are random, but
does not depend on that property. We therefore formulate the result for
deterministic operators. The relation to recent work \cite{CG03} will be
discussed below.

The Bulk is represented by the lattice $\Z^2 \ni x=(x_1,x_2)$ with Hamiltonian
$H_B = H_B^*$ on $\ell^2(\Z^2)$. We assume its matrix elements $H_B(x,x')$,
$x,x' \in \Z^2$, to be of short range in the sense that
\begin{equation}\label{eq:1}
    \sup_{x \in \Z^2} \sum_{x' \in \Z^2} \abs{H_B(x,x')} (\e^{\mu
    |x-x'|} - 1) \ =: \ C_1 \ < \ \infty
\end{equation}
for some $\mu > 0$, where $|x|=|x_1| + |x_2|$. Our hypothesis on the bounded
open interval $\Delta \subset \R$ is that for some $\nu \ge 0$
\begin{equation}\label{eq:2}
    \sup_{g \in B_1(\Delta)} \sum_{x,x' \in \Z^2}
    \abs{g(H_B)(x,x')} (1 + |x|)^{-\nu} \e^{\mu |x-x'|} \ =: \ C_2 \
    < \ \infty
\end{equation}
where $B_1(\Delta)$ denotes the set of Borel measurable functions $g$ which are
constant in $\set{\lambda | \lambda < \Delta}$ and in $\set{\lambda | \lambda >
\Delta}$  with $|g(x)| \le 1$ for every $x$.

In particular $C_2$ is a bound when $g$ is of the form $g_t(\lambda) = \e^{-\im
t \lambda} E_\Delta(\lambda)$ and the supremum is over $t \in \R$, which is a
statement of dynamical localization. By the RAGE theorem this implies that the
spectrum of $H_B$ is pure point in $\Delta$ (see \cite{KS80} or \cite[Theorem
9.21]{CFKS87} for details). We denote the corresponding eigen-projections by
$E_{\set{\lambda}}(H_B)$ for $\lambda \in \mathcal E_\Delta$, the set of
eigenvalues $\lambda \in \Delta$. We assume that no eigenvalue in $\mathcal
E_\Delta$ is infinitely degenerate,
\begin{equation}\label{eq:3}
    \dim E_{\set{\lambda}}(H_B) \ \ < \ \infty \; , \quad \lambda
    \in \mathcal E_\Delta \; .
\end{equation}
The validity of these assumptions is discussed below (but see also
\cite{AG98,ASFH01}).

The zero temperature bulk Hall conductance at Fermi energy $\lambda$ is defined
by the Kubo-St\v{r}eda formula \cite{ASS94}
\begin{equation}\label{eq:4}
    \sigma_B(\lambda) \ = \ - \im \tr P_\lambda
    \com{ \ \com{P_\lambda}{\Lambda_1}}{\com{P_\lambda}{\Lambda_2} \
    }\; ,
\end{equation}
where $P_\lambda = E_{(-\infty, \lambda)}(H_B)$ and $\Lambda_i(x)$ is the
characteristic function of $$\set{x=(x_1,x_2) \in \Z^2 \ | \ x_i < 0} \; .$$
Under the above assumptions $\sigma_B(\lambda)$ is well-defined for $\lambda
\in \Delta$, but independent thereof, i.e., it shows a plateau. (This result,
first proved in \cite{BvESB94}, is strengthened here in an appendix, since we
do not assume translation covariance or ergodicity of the Schr\"odinger
operator. We also show the integrality of $2 \pi \sigma_B$ therein, though it
is not needed in the sequel.) We remark that \eqref{eq:3} is essential for a
plateau: for the Landau Hamiltonian (though defined on the continuum rather
than on the lattice) eqs.~(\ref{eq:1}, \ref{eq:2}) hold if properly
interpreted, but \eqref{eq:3} fails in an interval containing a Landau level,
where indeed $\sigma_B(\lambda)$ jumps.

The sample with an Edge is modeled as a half-plane $\Z \times \Z_a$, where
$\Z_a = \set{n \in \Z \ | \ n \ge -a}$, with the height $-a$ of the edge
eventually tending to $-\infty$. The Hamiltonian $H_a = H_a^*$ on $\ell^2(\Z
\times \Z_a)$ is obtained by restriction of $H_B$ under some largely arbitrary
boundary condition.  More precisely, we assume that
\begin{equation}\label{eq:4a}
E_a = \J_a H_a - H_B \J_a \ : \ \ell^2(\Z \times \Z_a) \rightarrow \ell^2(\Z^2)
\end{equation}
satisfies
\begin{equation}\label{eq:5}
\sup_{x \in \Z^2} \sum_{x' \in \Z \times \Z_a} \abs{E_a(x,x')} \e^{\mu \left (
|x_2 +a| + |x_1 - x_1'| \right ) } \ \le \ C_3 \ < \ \infty \; ,
\end{equation}
where $\J_a: \ell^2(\Z \times \Z_a) \rightarrow \ell^2(\Z^2)$ denotes extension
by $0$. For instance with Dirichlet boundary conditions, $H_a = \J_a^* H_B
\J_a$, we have $E_a = (\J_a \J_a^* -1) H_B \J_a$, i.e.,
\begin{equation*}
    E_a(x,x') \ = \ \begin{cases}
      -H_B(x,x') \; , & x_2 < -a \; , \\
      0 \; , & x_2 \ge -a \; ,
    \end{cases}
\end{equation*}
whence \eqref{eq:5} follows from \eqref{eq:1}. We remark that \eq{eq:1} is
inherited by $H_a$ with a constant $C_1$ that is uniform in $a$, but not so for
\eq{eq:2} as a rule.

The definition of the edge Hall conductance requires some preparation.  The
current operator across the line $x_1 =0$ is $- \im \com{H_a}{\Lambda_1}$.
Matters are simpler if we temporarily assume that $\Delta$ is a gap for $H_B$,
i.e., if $\sigma(H_B) \cap \Delta = \emptyset$, in which case one may set
\cite{SBKR00}
\begin{equation}\label{eq:6}
    \sigma_E \ := \ - \im \tr \rho'(H_a) \com{H_a}{\Lambda_1} \; ,
\end{equation}
where $\rho \in C^\infty (\R)$ satisfies
\begin{equation}\label{eq:7}
    \rho(\lambda) \ = \ \begin{cases}
      1 \; , & \lambda < \Delta \; , \\
      0 \; , & \lambda > \Delta \; .
    \end{cases}
\end{equation}

The heuristic motivation for \eqref{eq:6} is as follows. We interpret
$\rho(H_a)$ as the 1-particle density matrix of a stationary quantum state.
Though some current is flowing near the edge we should discard it, as it is
supposed to be canceled by current flowing at an opposite edge located at $x_2
= + \infty$. If the chemical potential is now lowered by $\delta$ at the first
edge, but not at the second, a net current
\begin{equation*}
      I \ = \ - \im \tr \left ( \left ( \rho(H_a + \delta) -
    \rho(H_a) \right ) \com{H_a}{\Lambda_1} \right ) \
    = \ - \im \int_0^\delta \di t \tr \rho'(H_a - t)
    \com{H_a}{\Lambda_1}
\end{equation*}
is flowing. Since $\sigma_E$ is independent of $\rho$ as long as it conforms
with \eqref{eq:7}, see \cite{SBKR00} and Theorem \ref{thm} below, it is indeed
the conductance $\sigma_E = I/\delta$ for sufficiently small $\delta$.

The operator in \eqref{eq:6} is trace class essentially because $\im
\com{H}{\Lambda_1}$ is relevant only on (single-particle) states near $x_1 =0$,
and $\rho'(H_a)$ only near the edge $x_2 =-a$, so that the intersection of the
two strips is compact.  In the situation \eqref{eq:2} considered in this paper
the operator appearing in \eqref{eq:6} is not trace class, since the bulk
operator may have spectrum in $\Delta$, which can cause the above stated
property to fail for $\rho'(H_a)$. In search of a proper definition of
$\sigma_E$, we consider only the current flowing across the line $x_1=0$ within
a finite window $-a \le x_2 < 0$ next to the edge. This amounts to modifying
the current operator to be
\begin{equation}\label{eq:8}
    -\frac{\im}{2} \left ( \Lambda_2 \com{H_a}{\Lambda_1} +
    \com{H_a}{\Lambda_1} \Lambda_2 \right ) \ = \ - \frac{\im}{2}
    \set{\, \com{H_a}{\Lambda_1}  ,  \Lambda_2 } \; ,
\end{equation}
with which one may be tempted to use
\begin{equation}\label{eq:9}
    \lim_{a \rightarrow \infty} -\frac{\im}{2} \tr \rho'(H_a)
    \set{\, \com{H_a}{\Lambda_1}  ,  \Lambda_2 }
\end{equation}
as a definition for $\sigma_E$.  Though we show that this limit exists, it is
not the physically correct choice.  We may in fact expect that the dynamics of
$\e^{-\im t H_a}$ acting on states supported far away from the edge resembles
for quite some time the dynamics generated by $H_B$. Being bound states or,
more likely, resonances, such states may carry persistent currents (whence the
operator in \eqref{eq:6} is not trace class), but no or little \underline{net}
current across the line $x_1 = 0$. This cancelation is the rationale for
ignoring the part $x_2 \ge 0$ of the line $x_1 =0$ by means of the cutoff
$\Lambda_2$ in \eqref{eq:8}, however the cancelation is not achieved on states
located near the end point $x=(0,0)$. In the limit $a \rightarrow \infty$ we
pretend these states are bound, which
yields the contribution missed by \eqref{eq:9}:%
\begin{equation}\label{eq:11a}
    -\frac{\im}{2} \left ( \psi_\lambda,
    \set{\, \com{H_B}{\Lambda_1}  , 1-  \Lambda_2 }\psi_\lambda \right ) \ = \
    \Im \left ( \psi_\lambda, \Lambda_1 {H_B} \Lambda_2 \psi_\lambda\right ) \;
    ,
\end{equation}
from each bound state $\psi_\lambda$ of $H_B$, with corresponding energy
$\lambda \in \mathcal E_\Delta$. We incorporate them with weight
$\rho'(\lambda)$ in our definition of the edge conductance:
\begin{multline}\label{eq:11b}
    \sigma_E^{(1)} \ := \ \lim_{a \rightarrow \infty} -\frac{\im}{2} \tr
    \rho'(H_a) \set{\, \com{H_a}{\Lambda_1}  ,  \Lambda_2 } \\ + \sum_{\lambda \in
    \mathcal E_\Delta} \rho'(\lambda) \Im \tr E_{\set{\lambda}} \Lambda_1 H_B
    \Lambda_2 E_{\set{\lambda}} \; .
\end{multline}
We will show that the sum on the r.h.s.\ is absolutely convergent, and its
physical meaning will be further discussed at the end of the Introduction. We
will also show it to be non-zero on average for the Harper Hamiltonian with an
i.i.d.\ random potential in Theorem \ref{thm:harper}.

The terms of this sum involve $H_B$, though the few states for which they are
sizeable are supported near $x=(0,0)$ and hence far from the edge $x_2=-a$.
Since the mere appearance of $H_B$ in the definition of an edge property may be
objectionable, we present an alternative. The basic fact that the net current
of a bound state is zero,
\begin{equation}\label{eq:11}
    -\im \left ( \psi_\lambda , \com{H_B}{\Lambda_1} \psi_\lambda
    \right ) \ = \ 0 \; ,
\end{equation}
can be preserved by the regularization provided the spatial cutoff $\Lambda_2$
is time averaged. In fact, let
\begin{equation}\label{eq:12}
    A_{T,a}(X) \ = \ \frac{1}{T} \int_0^T \e^{\im H_a t} X
    \e^{-\im H_a t}\di t
\end{equation}
be the time average over $[0,T]$ of a (bounded) operator $X$ with respect to
the Heisenberg evolution generated by $H_a$, with $a$ finite or $a=B$. If a
limit $\Lambda_2^\infty = \lim_{T \rightarrow \infty} A_{T,B}(\Lambda_2)$ were
to exist, it would commute with $H_B$ so that
\begin{equation*}
    -\frac{\im}{2} \left ( \psi_\lambda, \set{\,
    \com{H_B}{\Lambda_1},\Lambda_2^\infty} \psi_\lambda \right ) \
    = \ 0 \; .
\end{equation*}
This motivates our second definition,
\begin{equation}\label{eq:13}
    \sigma_E^{(2)} \ :=  \lim_{T \rightarrow \infty} \lim_{a \rightarrow
    \infty} -\frac{\im}{2} \tr \rho'(H_a) \set{
    \,\com{H_a}{\Lambda_1},A_{T,a}(\Lambda_2)} \; .
\end{equation}

The two definitions allow for the following result.
\begin{theorem}\label{thm} Under the assumptions
(\ref{eq:1}, \ref{eq:2}, \ref{eq:3}, \ref{eq:5}, \ref{eq:7}) the sum in
(\ref{eq:11b}) is absolutely convergent, the limits there and in (\ref{eq:13})
exist, and
\begin{equation*}
    \sigma_E^{(1)} \ = \ \sigma_E^{(2)} \ = \  \sigma_B \; .
\end{equation*}
In particular (\ref{eq:11b}, \ref{eq:13}) depend neither on the choice of
$\rho$ nor on that of $E_a$.
\end{theorem}
\begin{remark}
  i.) The hypotheses (\ref{eq:1}, \ref{eq:2}) hold
  almost surely for ergodic Schr\"o\-dinger operators whose Green's
  function $G(x,x';z) = (H_B -z)^{-1}(x,x')$ satisfies a moment
  condition \cite{AM93} of the form
  \begin{equation}\label{eq:momentcondition}
   \sup_{E \in \Delta} \limsup_{\eta \rightarrow 0}
    \Ev{\abs{G(x,x';E + \im \eta)}^s} \ \le \ C \e^{- \mu |x-x'|}
  \end{equation}
  for some $s < 1$.  The implication is
  through the dynamical localization bound
  \begin{equation}\label{eq:meandynamicalloc}
    \Ev{\sup_{g \in B_1(\Delta)} \abs{g(H_B)(x,x')}} \ \le \ C
    \e^{-\mu |x-x'|} \; ,
  \end{equation}
  although \eqref{eq:2} has also been obtained by different means, e.g., \cite{GdB98}.
  The implication \eqref{eq:momentcondition} $\Rightarrow$ \eqref{eq:meandynamicalloc} was
  proved in \cite{Ai94} (see also \cite{AG98,dRJLS96,ASFH01}). The bound
  \eqref{eq:meandynamicalloc} may be better known for $\supp g
  \subset \Delta$, but is true as stated since it also holds
  \cite{BvESB94,AG98} for the projections $g(H_B) = P_\lambda$,
  $P_\lambda^\perp = 1 - P_\lambda$, ($\lambda \in \Delta$).

  ii.) Condition \eqref{eq:3}, in fact simple spectrum, follows form
   the arguments in \cite{Si94}, at least for operators with
   nearest neighbor hopping, $H_B(x,y)=0$ if $|x-y| >1$.

   iii.) When $\sigma(H_B) \cap \Delta = \emptyset$, the operator appearing
  in \eqref{eq:6} is known to be trace class. In
  this case, the conductance
  $\sigma_E^{(1)}=\sigma_E^{(2)}$ defined here
  coincides with $\sigma_E$ defined in \eqref{eq:6}. This statement
  follows from Theorem \ref{thm} and the known equality $\sigma_E = \sigma_B$
  \cite{SBKR00,EG02}, but can
  also be seen directly.  For completeness, we include a proof of this fact in
  Section 2 below.
\end{remark}

A point of view which combines both definitions of the edge conductance is
expressed by the following result.
\begin{theorem}\label{thm:instantaneous}
  Under the assumptions of Theorem \ref{thm},
  \begin{multline}\label{eq:19a}
    \lim_{a \rightarrow \infty} - \frac{\im}{2} \tr \rho'(H_a)
    \set{ \, \com{H_a}{\Lambda_1} \, , \, \Lambda_{2;a}(t)} \\
    = \ \sigma_B \ + \ \sum_{\lambda \in \mathcal E_\Delta}
    \rho'(\lambda) \Im \tr E_{\set{\lambda}}
    \com{H_B}{\Lambda_1} \e^{\im H_B t} \Lambda_2 \e^{-\im H_B t} E_{\set{\lambda}} \; ,
  \end{multline}
with $\Lambda_{2;a}(t) = \e^{\im H_a t} \Lambda_2 \e^{-\im H_a t}$.
\end{theorem}
In particular, this reduces to $\sigma_E^{(1)} = \sigma_B$ for $t=0$ by
(\ref{eq:11a}, \ref{eq:11b}).  On the other hand, $\sigma_E^{(2)} = \sigma_B$
results, as we will show, from the time average of \eqref{eq:19a}.

\vspace{.1in} A recent preprint \cite{CG03} contains results which are
topically related to but substantially different from those presented here. In
that work, two contiguous media are modeled by positing a potential of the form
$U(x_1,x_2)=V_0(x_2)\chi(x_2<0)+V(x_1,x_2)\chi(x_2\ge 0)$ (in our notation),
where $V_0$ is independent of $x_1$.  The role of $V$ is that of a bulk
potential, and that of $V_0$ as of a wall, provided it is large. The kinetic
term is given by the Landau Hamiltonian on the continuum $L^2(\R^2)$, whose
unperturbed spectrum is the familiar set $(2\N+1)B$, with $B$ the magnitude of
the constant magnetic field. A result is the following: if model (a), with
$V_0=0$, exhibits localization in $\Delta\subset[(2N-1)B, (2N+1)B]$ for some
positive integer $N$, and hence $\sigma_E=0$, then model (b), with $V_0(x_2)\ge
(2N+1)B$, has $2\pi\sigma_E=N$. The result is established by showing that the
difference between $2\pi\sigma_E$ in cases (b) and (a) is independent of $V,$
and equals $N$ if $V=0$, the two models then being solvable thanks to the
translation invariance w.r.t. $x_1$.

In comparison to our work, the following features may be noted:

i.) The localization assumption on the reference model (a) is made for a system
which has itself an interface. (Our eq. (1.2) concerns a bulk model serving as
reference.)

ii.) The validity of that assumption is limited to small $V$, because the
interface of (a) will otherwise produce extended edge states with energies in
$\Delta$. The result $\sigma_E=\sigma_B$ thus applies to perturbations of the
free Landau Hamiltonian of size $\lesssim B$. (Our comparison
$\sigma_E=\sigma_B$ does not require either side to be explicitly computable.)

iii.) The definition of $\sigma_E$ for (b)  depends on eigenstates in $\Delta$
of (a), like our $\sigma_E^{(1)}$, but not $\sigma_E^{(2)}$.

A model without bulk potential, but allowing interactions between diluted
particles, was studied from a related perspective in \cite{MMP88}.

\vspace{.1in} In (\ref{eq:11a}, \ref{eq:11b}) we argued that the limit
\eqref{eq:9} is not identical to $\sigma_B$.  To indeed prove this, we show
that the sum on the right hand side of \eqref{eq:11b} does not vanish for the
Harper Hamiltonian with i.i.d. Cauchy randomness on the diagonal.

The Harper Hamiltonian models the hopping of a tightly bound charged particle
in a uniform magnetic field. The hopping terms $H(x,x')$ are zero except for
nearest neighbor pairs, for which they are of modulus one,
\begin{equation}\label{eq:hopping}
    \abs{H(x,x')} \ = \ \begin{cases}
      0 \;, & |x-x'| \neq 1 \; , \\
      1 \;, & |x-x'| = 1 \; ,
    \end{cases}
\end{equation}
where the non-zero matrix elements are interpreted as
\begin{equation*}
    H(x,x') \ = \ \e^{\im \int_{x'}^x \mathbf A (y) \cdot \, \di^1 y} \;
    ,
\end{equation*}
with $\mathbf A$ the magnetic vector potential and the line integral computed
along the bond connecting $x,x'$. The magnetic flux through any region $D
\subset \R^2$ is
\begin{equation*}
    \int_D B(x) \di^2 x \ = \ \int_{\partial D} \mathbf A(y) \cdot
    \di^1 y \; ,
\end{equation*}
so, for a uniform field, the flux is proportional to the area
\begin{equation*}
    \int_{\partial D} \mathbf A(y) \cdot
    \di^1 y \ = \ \phi |D| \; .
\end{equation*}
Thus, we require that
\begin{equation}\label{eq:flux}
    H(x^{(1)},x^{(4)}) H(x^{(4)},x^{(3)})
    H(x^{(3)},x^{(2)})H (x^{(2)},x^{(1)}) \ = \ \e^{\im \int_{\partial P} \mathbf A(y) \cdot
    \di^1 y} \ = \ \e^{\im \phi}
\end{equation}
where $x^{(1)}, x^{(2)}, x^{(3)}, x^{(4)}$ are the vertices of a plaquette $P$,
listed in counter clockwise order and $\phi$ is the flux through any plaquette.

There are many choices of nearest neighbor hopping terms which satisfy
\eqref{eq:hopping} and \eqref{eq:flux}, all interrelated by gauge
transformations.  For our purposes, it suffices to fix a gauge and take
\begin{equation}\label{eq:Harper}
    H_\phi(x,x') \ := \ \begin{cases}
      1 \;, & x = x' \pm e_1 \; , \\
      \e^{\im \phi x_1} \; , & x = x' +e_2 \; , \\
      \e^{-\im \phi x_1} \; , & x = x' - e_2 \; ,
    \end{cases}
\end{equation}
with $e_1=(1,0)$ and $e_2=(0,1)$ the lattice generators. This choice of
$H_\phi$ comes from representing the constant field $B=\phi$ via the vector
potential $\mathbf A = \phi (0, x_1)$.  We note that the bulk and edge Hall
conductances are gauge invariant quantities, so Theorem \ref{thm:harper} stated
below holds for any other choice of $H_\phi$. We refer the reader to
ref.~\cite{OA01} and references therein for further discussion of the Harper
Hamiltonian.

To guarantee localized spectrum, we consider a bulk Hamiltonian which consists
of $H_\phi$ plus a diagonal random potential,
\begin{equation*}
    H_B \ = \ H_\phi + \alpha V
\end{equation*}
where $V \psi(x) \ = \ V(x) \psi(x)$ and $V(x)$, $x \in \Z^2$ are independent
identically distributed Cauchy random variables. Here $\alpha$ is a coupling
parameter (the ``disorder strength'') and ``Cauchy'' signifies that the
distribution of $v=V(x)$ is
\begin{equation*}
    \frac{1}{\pi} \frac{1}{1 + v^2} \di v \; .
\end{equation*}
We use Cauchy variables because it is possible to calculate certain quantities
explicitly for such variables: $\Ev{f(v)}=f(\im)$ for a function $f$ having a
bounded analytic continuation to the upper half plane.

It is clear that $H_B$ is short range, i.e., \eqref{eq:1} holds.  For
simplicity we consider $H_a$ which are defined via a non-random boundary
conditions, i.e., the operators $E_a$ appearing in \eqref{eq:4a} do not depend
on the random couplings $V(x)$. We then have  the following result.
\begin{theorem}\label{thm:harper} For $H_B, \ H_a$ as above,
there is $j_B \in C^\infty$ such that
\begin{equation}\label{eq:RadNik}
      \Ev{-\frac{\im}{2}  \lim_{a \rightarrow \infty} \tr \rho'(H_a)
    \set{\, \com{H_a}{\Lambda_1}, \Lambda_2 }}
    \ = \ -\int \rho'(\lambda) j_B(\lambda) \di \lambda \; ,
\end{equation}
whenever $\rho' \in C^\infty_0(\R)$. The expectation is well defined and may be
interchanged with the limit. Furthermore, $j_B(\lambda)$ has the following
asymptotic behavior
\begin{equation}\label{eq:asymptotic}
    j_B(\lambda) \ = \ -\frac{4 |\alpha|}{\pi}  \sin(\phi) ( \cos(\phi) +
    1) \lambda^{-5} \ + \ \mathcal O ( \lambda^{-6}) \; , \quad
    \abs{\lambda} \rightarrow \infty \; .
\end{equation}
\end{theorem}

The result is relevant in relation to \eqref{eq:11b} since it has in fact been
shown that \eqref{eq:2} holds for $H_B$ at large energies:
\begin{theorem*}[\cite{Ai94}]There is $E_0(\alpha)$ such that
\eqref{eq:meandynamicalloc} holds for $H_B$ and $\Delta = \Delta_\pm$ with
$\Delta_- =(-\infty, -E_0(\alpha)]$ and $\Delta_+ =[E_0(\alpha), \infty)$.
Hence \eqref{eq:2} holds almost surely.
\end{theorem*}
\begin{remark} i.) For any $\alpha \neq 0$ the spectrum of $H_B$ is (almost surely)
the entire real line, so the eigenvalues of $H_B$ in $\Delta_\pm$ make up a
(random) dense subset which we denote $\mathcal E_{\Delta_\pm}$. In fact, this
pure point spectrum is almost surely simple, as can be shown using the methods
in \cite{Si94}. ii.) For sufficiently large $\alpha$ we have $E_0(\alpha)=0$,
i.e., the spectrum is completely localized. iii.) Localization also holds
inside the spectral gaps of $H_\phi$, for small $\alpha$, via the methods in
\cite{Ai94,ASFH01}.
\end{remark}

The mentioned result implies $\sigma_B(\lambda)=0$ for $\lambda \in
\Delta_\pm$, because $\sigma_B$ is insensitive to $\lambda$ in that range and
$P_\lambda \rightarrow 1$ or $0$ as $\lambda \rightarrow \infty$ or $-\infty$,
respectively. Thus for $\rho$ as in \eqref{eq:7} with $\supp \rho' \subset
\Delta_\pm$ we have $\sigma_E^{(1)}=0$ by Theorem \ref{thm}. On the other hand,
for the first term on the r.h.s.\ of \eqref{eq:11b}, $J_B(\rho)$, we have by
Theorem \ref{thm:harper}
\begin{equation}\label{eq:28}
    \Ev{J_B(\rho)} \ = \   \frac{4 |\alpha|}{\pi}  \sin(\phi) ( \cos(\phi) +
    1)  \int_{|\lambda| \ge E_0(\alpha)} \rho'(\lambda) \lambda^{-5}\di
    \lambda \ + \ \mathcal{O} \left ( \lambda_0^{-6} \right )
\end{equation}
as $ \lambda_0 = \inf \set{|\lambda| | \lambda \in \supp \rho'}
    \rightarrow \infty$. Clearly the right hand side can be non-zero for
    appropriately chosen $\rho$, and the same then holds for the expectation of
    the last term in \eqref{eq:11b}.

\vspace{.1in} The definitions (\ref{eq:11b}, \ref{eq:13}) may be related,
heuristically, to concepts from classical electro-magnetism of material media
\cite{Ro73}.  There the macroscopic (or average) current is split as $\vec{j}_f
+
\partial \vec{P}/\partial t + \mathbf{rot} \, M$ into free, polarization, and
magnetization currents. (The magnetization $M$ is a scalar in two dimensions.)
The distinction depends on the existence of units (free electrons, atoms,
molecules, ...) each with conserved charge, whose current densities are
effectively of the form
\begin{equation}\label{eq:iii1}
\vec{j}(\vec{x},t) \ = \ q \dot{\vec{r}}(t) \delta (\vec{x} - \vec{r}(t)) +
\frac{\partial}{\partial t} \delta (\vec{x} - \vec{r}(t)) \vec{p}(t) +
\mathbf{rot} \left ( \delta (\vec{x} - \vec{r}(t)) m(t) \right ) \; ,
\end{equation}
where $q, \vec{p}(t), m(t)$ are the unit's charge and electric/magnetic moments
respectively. The macroscopic quantities emerge as a weak limit of the
microscopic ones
\begin{equation*}
\sum_{k} \delta (\vec{x} - \vec{r}_k(t)) \left \{
    \begin{array}{c}
        q_k \dot{\vec{r}}_k(t) \\
        \vec{p}_k(t) \\
        m_k(t)\\
    \end{array} \right \} \ \rightharpoonup \
    \left \{ \begin{array}{c}
        \vec{j}_f(\vec{x},t) \\
        \vec{P}(\vec{x},t) \\
        M(\vec{x},t)\\
    \end{array} \right \} \; ,
\end{equation*}
or more precisely after integration against compactly supported test
functions which vary slowly over the interatomic distance.  The microscopic
current across the portion $x_2 \le 0$ of the line $x_1=0$ is then
\begin{align}
I \ &= \ - \sum_k \int \di^2 \vec{x}
\Lambda'(x_1) \Lambda(x_2) \vec{j}_{k,1}(\vec{x},t) \notag \\
& \begin{aligned} = \ &- \int \di^2 \vec{x}
\Lambda'(x_1) \Lambda(x_2) \left [ \vec{j}_{f,1}(\vec{x},t)
+ \frac{\partial \vec{P}_1}{\partial t} (\vec{x},t) \right ]
\\ &+ \ \int \di^2 \vec{x}
\Lambda'(x_1) \Lambda'(x_2) M(\vec{x},t) \; .
\end{aligned}\label{eq:iii2}
\end{align}
The derivation assumes that $\Lambda$ is smooth over interatomic distances.
The last term in \eqref{eq:iii2} comes from the corresponding term in
\eqref{eq:iii1}, which is $\partial_2 \delta(\vec{x} - \vec{r}_k(t))
m_k(t)$.  It cannot be replaced by adding $(\mathbf{rot} M)_1 = \partial_2 M$
within the square brackets, which would correspond to the macroscopic current.  In
fact, it differs from that by a boundary term, which would vanish if
$\Lambda(x_2)$ were compactly supported. Let now the macroscopic fields be
stationary and slowly varying on the scale of $\Lambda'$.  In the QHE we
expect that the (free) edge currents are located near the edge, so that
\eqref{eq:iii2} becomes
\begin{equation*}
I \ =  \int_{x_1 = 0 } \di x_2 \vec{j}_{f,1} (\vec{x}) \ + \ M(0) \; .
\end{equation*}
When $M(0)$ is subtracted from the l.h.s., we obtain an expression for the
edge current, which is the role of the second term in \eqref{eq:11b}.  In
this analogy the definition \eqref{eq:13} corresponds to replacing
$\Lambda(x_2)$ in the first line of \eqref{eq:iii2} by
$\Lambda(\vec{e}_2 \cdot \vec{r}_{k,T})$ where $\vec{r}_{k,T}$ is the time average of
$\vec{r}_k(t)$. Then the last term no longer arises.

The above discussion neglects the weighting $\rho'(\lambda)$ of energies in
\eqref{eq:11b}. This will be remedied in the following heuristic argument in
support of $\sigma_B = \sigma_E$.  In a finite sample of volume $V$ the
St\v{r}eda relation \cite{St82} asserts
\begin{equation}\label{eq:final1}
\frac{\partial N}{\partial \phi} \ \cong \ \sigma_B V \; ,
\end{equation}
where $N$ is the total charge of carriers, i.e., $N= \tr \rho(H_V)$ in the
situation considered here. For the total magnetization $M$ we have
\begin{equation}\label{eq:final2}
- \frac{\partial M}{\partial \mu} \ = \ \tr \rho'(H_V) \frac{\partial
H_V}{\partial \phi} \; ,
\end{equation}
where $\mu$ is the chemical potential, as can be seen from the Maxwell relation
\cite{GA03}
\begin{equation}\label{eq:final3}
    - \frac{\partial M}{\partial \mu} \ = \ \frac{\partial N}{\partial \phi} \;
    .
\end{equation}
To compute $\partial H_V/\partial \phi$ we use a gauge equivalent to
\eqref{eq:Harper}, with trivial phases along bonds in direction $\vec{e}_2$,
and obtain for \eqref{eq:final2}
\begin{equation*}
    -\frac{1}{V} \frac{\partial M}{\partial \mu} \ = \ \frac{1}{V}
    \frac{\im}{2} \tr \rho'(H_V) \set{ \, \com{H_V}{X_1}\, , \, X_2} \; .
\end{equation*}
By (\ref{eq:final1}, \ref{eq:final3}) this quantity is formally $\sigma_B$. To
relate it to $\sigma_E^{(1)}$ it should be noted that the total magnetization
is \underline{not} the integral of the bulk magnetization, even in the
thermodynamic limit.  For instance, for classical, spinless particles $M$
vanishes \cite{vLe21}, but consists \cite{Pe79} of a diamagnetic, bulk
contribution and a an opposite contribution from states close to the edge.
These two contributions (in reverse order) may be identified in the quantum
mechanical context with the two terms of \eqref{eq:11b}. In this example, the
expected edge term is negative for $\phi > 0$. This should also emerge from
\eqref{eq:28} when $\sup \supp \rho' \rightarrow - \infty$, and it does if one
also takes into account that $-H$ is the counterpart to the continuum
Hamiltonian.

In Section \ref{sec:outline} we will present the main steps in the proof of
Theorems \ref{thm} and \ref{thm:instantaneous}, with details supplied in
Section \ref{sec:proofs}. The proof of Theorem \ref{thm:harper} will be given
in Section \ref{sec:harper}. The appendix is about properties of $\sigma_B$.

\begin{acknowledgements} We thank M. Aizenman, Y. Avron,
J. Bellissard, J.-M. Combes,  J. Fr\"ohlich, F. Germinet, and H. Schulz-Baldes
for useful discussions.
\end{acknowledgements}

\section{Outline of the proof}\label{sec:outline}
A reasonable first step is to make sure that the traces in (\ref{eq:11b},
\ref{eq:13}) are well-defined. We will show this for
\begin{equation}\label{eq:16}
    \sigma_E(a,t) \ := \ - \im \tr \rho'(H_a)
    \com{H_a}{\Lambda_1}\Lambda_{2;a}(t) \; ,
\end{equation}
with $\Lambda_{2;a}(t)  =  \e^{\im t H_a} \Lambda_2 \e^{-\im t H_a}$, by
proving that $\im \com{H_a}{\Lambda_1} \Lambda_{2;a}(t) \in \I_1$ in
Lemma~\ref{lem:traceclass}. Here, $\I_1$ denotes the ideal of trace class
operators, and we denote the trace norm by $\norm{\cdot}_1$. Then
\begin{equation*}
\begin{split}
    \overline{\sigma_E(a,t)} \ &= \ -\im \tr \Lambda_{2;a}(t)
    \com{H_a}{\Lambda_1}\rho'(H_a) \\
    &= \ - \im \tr \rho'(H_a) \Lambda_{2;a}(t) \com{H_a}{\Lambda_1} \; ,
\end{split}
\end{equation*}
where we used that
\begin{equation}\label{eq:44a}
 \tr AB \ = \ \tr BA \text{ if } AB \; , \ BA \ \in \ \I_1 \; ,
\end{equation}
e.g., \cite[Corollary 3.8]{Si79}. The definition (\ref{eq:13})
then reads
\begin{equation}\label{eq:17}
    \sigma_E^{(2)} \ = \ \lim_{T \rightarrow \infty} \lim_{a \rightarrow
    \infty} \frac{1}{T} \int_0^T \di t \Re \sigma_E(a,t) \; .
\end{equation}

By the argument given in the Introduction, the trace norm of the operator in
\eqref{eq:16} diverges as $a \rightarrow \infty$.  To see that its trace
nevertheless converges we subtract from it an operator $Z(a,t) \in \I_1$, to be
specified below, with $\tr Z(a,t)  =  0 $, implying
\begin{equation}\label{eq:19}
    \sigma_E(a,t) \ = \ - \im \tr \left ( \rho'(H_a)
    \com{H_a}{\Lambda_1}\Lambda_{2;a}(t) - Z(a,t) \right ) \; .
\end{equation}
The idea, of course, is to choose $Z(a,t)$ so that
\begin{equation}\label{eq:20}
    \sup_{a} \norm{ \rho'(H_a) \com{H_a}{\Lambda_1} \Lambda_{2;a}(t) -
    Z(a,t) }_1 \ < \ \infty \; .
\end{equation}

An operator of zero trace is $\com{\rho(H_a)}{\Lambda_1} \Lambda_2$; it is
trace class (see Lemma \ref{lem:traceclass}) and its trace, computed in the
position basis, is seen to vanish. Though it does not quite suffice for
\eqref{eq:20}, we consider it since $\com{\rho(H_a)}{\Lambda_1}$ and
$\rho'(H_a) \com{H_a}{\Lambda_1}$ are closely related: From the
Helffer-Sj\"ostrand representations (see Section 3 for details)
\begin{subequations}\label{eq:20a}
\begin{align}
  \rho(H_a) \ &= \ \frac{1}{2\pi} \int \di m(z) \partial_{\bar z}
  \rho(z) R(z)  \label{eq:20a1} \\
  \rho'(H_a) \ &= \ -\frac{1}{2\pi} \int \di m(z) \partial_{\bar z}
  \rho(z) R(z)^2 \; , \label{eq:20a2}
\end{align}
\end{subequations}
with $R(z) = (H_a-z)^{-1}$, we obtain
\begin{subequations}\label{eq:21}
\begin{align}\label{eq:21a}
  \com{\rho(H_a)}{\Lambda_1} \ &= \ -\frac{1}{2\pi} \int \di m(z)
  \partial_{\bar z} \rho(z) R(z) \com{H_a}{\Lambda_1} R(z) \\
  \rho'(H_a)\com{H_a}{\Lambda_1} \ &= \ -\frac{1}{2\pi} \int \di m(z)
  \partial_{\bar z} \rho(z) R(z)^2\com{H_a}{\Lambda_1} \; .\label{eq:21b}
\end{align}
\end{subequations}
The two expressions, multiplied from the right by $\Lambda_2$, respectively by
$\Lambda_{2;a}(t)$ as in \eqref{eq:16}, would have an even more similar
structure if in the second a resolvent could be moved to the right. This can be
achieved under the trace by setting
\begin{multline}\label{eq:21c}
  Z(a,t) \ = \ \com{\rho(H_a)}{\Lambda_1} \Lambda_2 \\
  - \frac{1}{2\pi} \int \di m(z) \partial_{\bar z} \rho(z) R(z)
  \left ( R(z) \com{H_a}{\Lambda_1} \Lambda_{2;a}(t) -
  \com{H_a}{\Lambda_1} \Lambda_{2;a}(t) R(z) \right ) \; ,
\end{multline}
for which $\tr Z(a,t)  =  0$. Then \eqref{eq:19} reads $
    \sigma_E(a,t) \ = \ \tr \Sigma_a(t)$
with
\begin{align}\label{eq:22}
 \im \Sigma_a(t) \  :&= \
  \overbrace{- \com{\rho(H_a)}{\Lambda_1} \Lambda_2}^{\im \widetilde \Sigma'_a}
  \\ & \quad   +  \
  \underbrace{-\frac{1}{2 \pi} \int \di m(z) \partial_{\bar z} \rho(z) R(z)
  \com{H_a}{\Lambda_1}\Lambda_{2;a}(t) R(z)}_{\im \widetilde \Sigma''_a(t)}
  \notag \\ \label{eq:23}
    &= \
  \overbrace{\com{\rho(H_a)}{\Lambda_1} \left ( \Lambda_{2;a}(t) - \Lambda_2
  \right )}^{\im \Sigma_a'(t)} \\  \notag & \quad  + \
  \underbrace{-\frac{1}{2 \pi} \int \di m(z) \partial_{\bar z} \rho(z) R(z)
        \com{H_a}{\Lambda_1}R(z) \com{H_a}{\Lambda_{2;a}(t)} R(z)}_{\im \Sigma_a''(t)}
         ,
\end{align}
where, to obtain the last expression, \eqref{eq:21a} multiplied by
$\Lambda_{2;a}(t)$ has been added and subtracted, and
$\com{R(z)}{\Lambda_{2;a}(t)} = - R(z) \com{H_a}{\Lambda_{2;a}(t)} R(z)$ has
been used. We remark that equality of \eqref{eq:22} and \eqref{eq:23} also
holds for $H_B$, i.e., if we replace $H_a$ by $H_B$ and $\Lambda_{2;a}(t)$ by $
\Lambda_{2;B}(t) = \e^{\im H_B t} \Lambda_2 \e^{-\im H_B t}$, and set $R(z) =
(H_B -z )^{-1}$.

We will show
\begin{equation}\label{eq:24}
    \sigma_E(a,t) \ = \ \tr \Sigma_a(t) \ \xrightarrow[a \rightarrow
    \infty]{} \ \tr \Sigma_B(t)
\end{equation}
and, incidentally, \eqref{eq:20} by establishing:
\begin{lemma}\label{lem:tracelimit}
  Under assumptions (\ref{eq:1}, \ref{eq:5}), but without making use of
  (\ref{eq:2}, \ref{eq:3}, \ref{eq:7}), we have for $\rho' \in C_0^\infty(\R)$
  \begin{align}\label{eq:25}
    \norm{\J_a \Sigma_a'(t) \J_a^* - \Sigma_B'(t)}_1 \ & \xrightarrow[a \rightarrow
    \infty]{}  \ 0 \; , \\ \label{eq:26}
    \norm{\J_a \Sigma_a''(t) \J_a^* - \Sigma_B''(t)}_1 \ & \xrightarrow[a \rightarrow
    \infty]{}  \ 0 \;
  \end{align}
  uniformly for $t$ in a compact interval.
\end{lemma}
Note that the replacement $A \mapsto \J_a A \J^*_a$ simply extends by zero an
operator on $\ell^2(\Z \times \Z_a)$ to one on $\ell^2(\Z^2)$. In particular
$\norm{\J_a A \J^*_a}_1 = \norm{A}_1$ and $\tr \J_a A \J^*_a = \tr A$.

{For the rest of this section} on we shall only be concerned with Bulk
quantities like $\tr \Sigma_B(t)$. By (\ref{eq:16}, \ref{eq:24}), the
statements to be proven are
\begin{equation*}
    \Re \tr\Sigma_B(t) \ = \ \sigma_B \ + \ \sum_{\lambda \in \mathcal E_\Delta}
    \rho'(\lambda) \Im \tr E_{\set{\lambda}}
    \com{H_B}{\Lambda_1} \e^{\im H_B t} \Lambda_2 \e^{-\im H_B t} E_{\set{\lambda}}
\end{equation*}
for Theorem 2 and part of Theorem 1, and
\begin{equation}\label{eq:2.14a}
    \lim_{T \rightarrow \infty} \frac{1}{T} \int_0^T \tr \Sigma_B(t) \di t  \ = \ \sigma_B
\end{equation}
for the other part, where actually the real part of the l.h.s.\ would suffice.
It may be noted that the $\rho$'s allowed by \eqref{eq:7} form an affine space
and that $\Sigma_B(t)$, like $\sigma_E^{(1)}$, $\sigma_E^{(2)}$, is affine in
$\rho$.  The relation to $\sigma_B$ will be made through the following
decomposition, which exhibits the same property for this quantity.
\begin{lemma}\label{lem:sigmaBrep}
Let $\Delta \subset \R$ be as in Theorem \ref{thm} and let $E_-$, $E_+$ be the
spectral projections for $H_B$ onto $\set{\lambda \, | \, \lambda < \Delta}$,
resp. $\set{\lambda \, | \, \lambda > \Delta}$. Then, for $\lambda_0 \in
\Delta$
\begin{equation}\label{eq:29}
      \sigma_B(\lambda_0) \ = \ \ \im \tr E_-
      \com{P_{\lambda_0}}{\Lambda_1} \Lambda_2 E_-  +
      \im  \tr E_+ \com{P_{\lambda_0}}{\Lambda_1} \Lambda_2 E_+ +
      \im \tr E_\Delta T_{\lambda_0} E_\Delta \; ,
\end{equation}
where $P_{\lambda_0} = E_{(-\infty,\lambda_0)}$,
\begin{equation}\label{eq:29a}
T_{\lambda_0} \ = \ P_{\lambda_0} \Lambda_1 P_{\lambda_0}^\perp \Lambda_2
P_{\lambda_0} - P_{\lambda_0}^\perp \Lambda_1 P_{\lambda_0} \Lambda_2
P_{\lambda_0}^\perp
\end{equation}
and the traces are well defined.  Moreover, the last term in \eqref{eq:29} can
be further decomposed as
\begin{equation}\label{eq:29b}
\im \tr E_\Delta T_{\lambda_0} E_\Delta  \ = \ \sum_{\lambda \in \mathcal
E_\Delta} \im \tr E_{\{\lambda\}}
      \com{P_{\lambda_0}}{\Lambda_1} \Lambda_2 E_{\{\lambda\}} \; ,
\end{equation}
with absolutely convergent sum.
\end{lemma}

Since $\sigma_B$ is independent of $\lambda_0 \in \Delta$, (\ref{eq:29}) with
the last term replaced by the r.h.s.\ of (\ref{eq:29b}) also holds if
$P_{\lambda_0}$ is replaced by $\rho$ satisfying \eqref{eq:7}, since $\rho(H_B)
= - \int \di \lambda_0 \rho'(\lambda_0) P_{\lambda_0}$.  The proof of Lemma
\ref{lem:sigmaBrep}, which is given in Section \ref{sec:proofs}, makes use of
\begin{equation}\label{eq:30}
    1 \ = \ E_- \ + \ E_+ \ + \ E_\Delta \; , \quad E_\Delta =
    \sum_{\lambda \in \mathcal
    E_{\Delta}} E_{\{\lambda\}} \; ,
\end{equation}
where the sum is strongly convergent. Using this decomposition on $\Sigma_B(t)
\in \I_1$ we obtain
\begin{multline}\label{eq:30a}
    \tr \Sigma_B(t) \ = \ \tr E_- \left (\widetilde
      \Sigma_B' + \widetilde \Sigma_B''(t) \right ) E_- + \tr E_+ \left (\widetilde
      \Sigma_B' + \widetilde \Sigma_B''(t) \right ) E_+ \\
        + \tr E_\Delta \Sigma_B(t) E_\Delta \; .
\end{multline}
Though the two contributions \eqref{eq:22} to $\Sigma_B(t)$ are not separately
trace class, they become so in \eqref{eq:30a}. In fact, those of $E_\pm
\widetilde \Sigma'_BE_\pm$ also appear in \eqref{eq:29}, and $E_\pm \widetilde
\Sigma_B''(t) E_\pm$ vanish by integration by parts since $E_\pm R(z)$ and
$R(z) E_\pm $ are analytic on the support of $\rho(z)$ or of $\rho(z) -1$. We
thus find that
\begin{equation}\label{eq:30b}
    \tr \Sigma_B(t) \ = \ \sigma_B + \im \int \di \lambda_0 \rho'(\lambda_0)
    \tr E_\Delta T_{\lambda_0} E_\Delta + \tr E_\Delta \Sigma_B(t) E_\Delta \;
    .
\end{equation}

At this point the analysis of the last term splits into two tracks with the
purpose of showing $\sigma_E^{(1)} = \sigma_B$, resp.\ $\sigma_E^{(2)} =
\sigma_B$.

\subsection{Track 1}
We decompose the projection $E_\Delta$ into its atoms as in \eqref{eq:30},
which by \begin{equation}\label{eq:95}
    X_n \xrightarrow[]{s} 0 \, , \ Y \in \I_1 \ \implies \
    \norm{X_n Y}_1 \rightarrow 0 \, , \ \norm{Y X_n^*}_1
    \rightarrow 0 \;
\end{equation}
yields a trace class norm convergent sum for
$E_\Delta \Sigma_B(t) E_\Delta$. Thus
\begin{equation*}
    \tr E_\Delta \Sigma_B(t) E_\Delta \ = \ \sum_{\lambda \in \mathcal E_\Delta}
    \tr E_{\set{\lambda}} \left ( \widetilde \Sigma_B' + \widetilde
    \Sigma_B''(t) \right ) E_{\set{\lambda}} \; .
\end{equation*}
Again, the contributions $E_{\set{\lambda}} \widetilde \Sigma_B'
E_{\set{\lambda}}$ are themselves trace class as they match those of
\eqref{eq:29b}, canceling the second term of \eqref{eq:30b}. We conclude that
\begin{equation}\label{eq:31}
    \begin{split}
      \tr &\Sigma_B(t) \\ & = \ \sigma_B \ + \ \frac{\im}{2 \pi}
      \sum_{\lambda \in \mathcal E_\Delta} \int \di m(z)
      \partial_{\bar z} \rho(z) \tr E_{\{\lambda\}} R(z)
      \com{H_B}{\Lambda_1} \Lambda_{2;B}(t) R(z) E_{\{\lambda\}} \\
      & = \ \sigma_B \ - \ \im \sum_{\lambda \in \mathcal
      E_\Delta}\rho'(\lambda) \tr E_{\{\lambda\}} \e^{-\im H_B t}
      \com{H_B}{\Lambda_1} \e^{\im H_B t} \Lambda_2
      E_{\{\lambda\}}\; ,
    \end{split}
\end{equation}
where we used that $f(H_B) E_{\{ \lambda\}} = f(\lambda) E_{\{\lambda\}}$. By
its derivation this sum is absolutely convergent for each $t$. This proves
Thm.~\ref{thm:instantaneous} and hence $\sigma_E^{(1)} = \sigma_B$.

\subsection{Track 2}
Here we do not decompose $E_\Delta$, but use \eqref{eq:23} whose two terms are
separately trace class,
\begin{equation*}
    \tr E_\Delta \Sigma_B(t) E_\Delta \ = \ \tr E_\Delta \Sigma_B'(t) E_\Delta
    + \tr E_\Delta \Sigma_B''(t) E_\Delta \; .
\end{equation*}
\begin{lemma}\label{lem:track2}
For $\Delta \subset \R$ as in Theorem \ref{thm} we have
\begin{equation}\label{eq:31b}
    \frac{1}{T} \int_0^T \tr E_\Delta \Sigma_B''(t) E_\Delta \, \di t \ \xrightarrow[T
    \rightarrow \infty]{} \ 0 \; ,
\end{equation}
and
\begin{equation}\label{eq:31a}
    - \im \tr E_\Delta \com{P_{\lambda_0}}{\Lambda_1} \left (
    A_{T,B}(\Lambda_2) - \Lambda_2 \right ) E_\Delta \ \xrightarrow[T
    \rightarrow \infty]{} \ \im \tr E_\Delta T_{\lambda_0} E_\Delta
\end{equation}
for $\lambda_0 \in \Delta$, the expression on the l.h.s.\ being uniformly
bounded in $\lambda_0 \in \Delta$, $T>0$.
\end{lemma}

By dominated convergence \eqref{eq:31a} implies
\begin{equation*}
    \frac{1}{T} \int_0^T \tr E_\Delta \Sigma_B'(t) E_\Delta \, \di t \ \xrightarrow[T
    \rightarrow \infty]{} \ - \im \int \di \lambda_0 \rho'(\lambda_0) \tr
    E_\Delta T_{\lambda_0} E_\Delta \; .
\end{equation*}
Together with (\ref{eq:30b}, \ref{eq:31b}), this proves \eqref{eq:2.14a} and
hence $\sigma_E^{(2)} = \sigma_B$.

\subsection{Alternate Track 2}We now show
that the last result can also be inferred from \eqref{eq:31}, at least if
assumption \eqref{eq:3} is strengthened to a uniform upper bound on the
degeneracies:
\begin{equation}\label{eq:31c}
    \dim E_{\set{\lambda}} (H_B) \ \le \ C_4 < \infty \; , \quad \lambda \in
    \mathcal E_\Delta \; .
\end{equation}
Then, the sum \eqref{eq:31} is uniformly convergent in $t \in \R$, as stated in
\begin{lemma}\label{lem:uniformintbound}
  Assuming (\ref{eq:1}, \ref{eq:2}, \ref{eq:31c}), we have
  \begin{equation}\label{eq:32}
    \sum_{\lambda \in \mathcal E_\Delta} \sup_{t \in \R} \abs{
    \tr  E_{\set{\lambda}}
    \e^{-\im H_B t} \com{H_B}{\Lambda_1} \e^{\im
    H_B t} \Lambda_2 E_{\set{\lambda}} } \ < \ \infty \; .
\end{equation}
\end{lemma}
In order to prove \eqref{eq:2.14a}, it suffices in view of \eqref{eq:32} to
show
\begin{equation}\label{eq:33}
    \lim_{T \rightarrow \infty} \frac{1}{T} \int_0^T \tr
    E_{\set{\lambda}} \e^{-\im H_B t} \com{H_B}{\Lambda_1} \e^{\im
    H_B t} \Lambda_2 E_{\set{\lambda}} \ = \ 0
\end{equation}
for each $\lambda \in \mathcal E_{\Delta}$. Because
\begin{equation}\label{eq:33a}
    \tr
    E_{\set{\lambda}} \e^{-\im H_B t} \com{H_B}{\Lambda_1} \e^{\im
    H_B t} \Lambda_2  E_{\set{\lambda}} \ = \ \im \frac{\di}{\di t}
    \ \tr
    E_{\set{\lambda}} \e^{-\im H_B t} \Lambda_1 \e^{\im
    H_B t} \Lambda_2  E_{\set{\lambda}} \; ,
\end{equation}
the expression under the limit is just
\begin{equation}\label{eq:33b}
    \frac{\im}{T} \left [ \tr E_{\set{\lambda}} \e^{-\im H_B t} \Lambda_1 \e^{\im
    H_B t} \Lambda_2   -
    \tr
    E_{\set{\lambda}}  \Lambda_1  \Lambda_2
    \right ] \; .
\end{equation}
Since each term inside the square brackets is bounded by $C_4 < \infty$,
\eq{eq:33} follows. This concludes the alternate proof of $\sigma_E^{(2)} =
\sigma_B$.

\subsection{Edge conductance in a spectral gap} We conclude this section by showing as
mentioned above in the remark following Theorem \ref{thm} that
\begin{equation*}
    -\im \tr \rho'(H_a) \com{H_a}{\Lambda_1} \ = \ \sigma_B
\end{equation*}
if $\sigma(H_B) \cap \Delta = \emptyset$.  By translation invariance of
$\sigma_B$, see Lemma \ref{lem:translationinvariance} below, it suffices to
show this for $a=0$, in which case we drop the subscript $a$ of the edge
Hamiltonian.  It has been shown in (A.8) of \cite{EG02} that $\rho'(H)
\com{H}{\Lambda_1} \in \I_1$. Since $\Lambda_{2,a} := \Lambda_2(\cdot - a)
\xrightarrow[]{s} 1$ as $a \rightarrow \infty$, we have by \eqref{eq:95}
\begin{multline}\label{eq:gap1}
- \im \tr \rho'(H) \com{H}{\Lambda_1} \ = \ - \im \lim_{a \rightarrow
\infty} \tr \rho'(H) \com{H}{\Lambda_1} \Lambda_{2,a}  \\
= \ - \im \lim_{a \rightarrow \infty} \tr \rho'(H^a) \com{H^a}{\Lambda_1}
\Lambda_2 \; .
\end{multline}
Here $H^a$ is the operator on $\ell^2(\Z \times \Z_a)$ obtained from $H$ by a
shift $(0,-a)$; it is not the restriction to $\Z \times \Z_a$ of a fixed Bulk
Hamiltonian $H_B$, as $H_a$ was, but instead of an equally shifted one,
$H_B^a$. The estimates (\ref{eq:1}, \ref{eq:5}) therefore still apply, which is
all that matters for (\ref{eq:25}, \ref{eq:26}). The r.h.s.\ of \eqref{eq:gap1}
thus equals $\lim_{a \rightarrow \infty} \tr \Sigma_B^a(0)$, where
$\Sigma_B^a(t)$ pertains to $H_B^a$. Since the sum in \eqref{eq:31} vanishes,
$\tr \Sigma_B^a(t)  = \sigma_B^a$, which is independent of $a$.

\section{Details of the Proof}\label{sec:proofs}
We give some details about the Helffer-Sj\"ostrand representations
(\ref{eq:20a}). The integral is over $z=x+ \im y \in \C$ with measure $\di m(z)
= \di x \di y$, $\partial_{\bar z} = \partial_x + \im \partial_y$, and
$\rho(z)$ is a quasi-analytic extension of $\rho(x)$ which, see \cite{HS00},
for given $n$ can be chosen so that
\begin{equation}\label{eq:37}
    \int \di m(z) \abs{\partial_{\bar z} \rho(z) } |y|^{-p-1} \
    \le \ C \sum_{k=0}^{n+2} \norm{\rho^{(k)}}_{k-p-1}
\end{equation}
for $p=1,...,n$, provided the appearing norms $
    \norm{f}_k = \int \di x (1+x^2)^{\frac{k}{2}} |f(x)|$
are finite. This is the case for $\rho$ with $\rho' \in C_0^\infty(\R)$. For
$p=1$ this shows that \eqref{eq:20a2} is norm convergent. The integral
\eqref{eq:20a1}, which would correspond to the case $p=0$, is nevertheless a
strongly convergent improper integral, see e.g., (A.12) of \cite{EG02}.

A further preliminary is the Combes-Thomas bound \cite{CT73}
\begin{equation}\label{eq:38}
\norm{\e^{\delta \ell(x) } R_a(z) \e^{-\delta \ell(x)}} \ \le \ \frac{C}{|\Im
z|}  \; ,
\end{equation}
where $\delta$ can be chosen as
\begin{equation}\label{eq:39}
    \delta^{-1} \ = \ C \left ( 1 + |\Im z |^{-1} \right )
\end{equation}
for some (large) $C >0$ and $\ell(x)$ is any Lipschitz function on $\Z^2$ with
\begin{equation}\label{eq:39a}
    \abs{\ell(x) - \ell(y)} \ \le \ | x- y| \;
\end{equation}
(see e.g. \cite[Appendix D]{AG98} for details).

\begin{lemma}\label{lem:traceclass}
  We have
  \begin{equation}\label{eq:40}
    \com{H_a}{\Lambda_1} \Lambda_{2;a}(t) \ \in \ \I_1 \; ,
  \end{equation}
  and for $\rho \in C^\infty(\R)$ with $\supp \rho'$ compact also
  \begin{equation}\label{eq:41}
    \com{\rho(H_a)}{\Lambda_1} \Lambda_{2;a}(t) \ \in \ \I_1 \; .
  \end{equation}
  In particular, $Z(a,t)$ as given in \eqref{eq:21c} is trace
  class.
\end{lemma}
\begin{proof}
  We first prove the finite propagation speed estimate (see
  \cite{AES04} and \cite{LR72}):
  \begin{quotation}\noindent Let $\mu >0$ be as in
  \eqref{eq:1}. Then, for $0 \le \delta \le \mu$ and $\ell$ as \eqref{eq:39a},
  \begin{equation}\label{eq:42}
    \norm{\e^{\delta \ell(x)} \e^{\im H_a t} \e^{-\delta \ell(x)}}
    \ \le \ \e^{C |t|}
  \end{equation}
  for some $C < \infty$.
  \end{quotation}
  Indeed, let $A(t)=\e^{\delta \ell(x)} \e^{\im H_a t}
  \e^{-\delta \ell(x)}$.
  \begin{equation*}
  \frac{\di}{\di t} A(t)^* A(t) \ = \
  \frac{\di}{\di t}
    \e^{-\delta \ell(x)} \e^{-\im H_a t} \e^{2 \delta \ell(x)} \e^{\im H_a
    t} \e^{-\delta \ell(x)} \ = \  A(t)^* B A(t) \; ,
  \end{equation*}
  where $B = - \im \e^{-\delta \ell(x)} \com{H_a}{\e^{2 \delta \ell(x)}}
  \e^{-2 \delta \ell(x)}$ has matrix elements
  \begin{equation*}
    \im B(x,x') \ = \ H_a(x,x') (\e^{\delta (\ell(x')-\ell(x))} -
    \e^{\delta(\ell(x) -\ell(x'))} ).
  \end{equation*}
  By \eqref{eq:1} which, as remarked in the Introduction,
  is inherited by $H_a$, and by Holmgren's bound
  \begin{equation}\label{eq:43}
    \norm{B} \ \le \ \max \left ( \sup_x \sum_{x'} |B(x,x')| \, ,
    \ \sup_{x'} \sum_x |B(x,x')|
    \right ) \; ,
  \end{equation}
  we have $2 C := \norm{B} < \infty$ and hence
  \begin{equation*}
    \norm{A(t)}^2 \ = \ \norm{A(t)^* A(t)} \ \le \ \e^{2 C |t|} \; .
  \end{equation*}

  We factorize
  \begin{equation*}
    \com{H_a}{\Lambda_1} \Lambda_{2;a}(t) \ = \
    \com{H_a}{\Lambda_1}\e^{\delta |x_1|} \cdot \e^{-\delta
    |x_1|}\e^{-\delta |x_2|} \cdot \e^{\delta |x_2|} \Lambda_{2;a}(t) \; ,
  \end{equation*}
  and note that
  \begin{equation}\label{eq:44}
    \norm{\e^{-\delta |x_1|} \e^{-\delta |x_2|}}_1 \ \le \ C
    \delta^{-2} \; ,
  \end{equation}
  since this is a summable function of $(x_1,x_2) \in \Z^2$. It is therefore
  enough for \eqref{eq:40} to show
  \begin{gather}\label{eq:45}
    \norm{\com{H_a}{\Lambda_1}\e^{\delta |x_1|}} \ \le \ C \; , \\
    \label{eq:46}
    \norm{\e^{\delta |x_2|} \Lambda_{2;a}(t)} \ \le \ C\e^{\delta a}
  \end{gather}
  for small $\delta$, where the first estimate also holds for $a=B$.
  Indeed, the first operator has matrix
  elements
  \begin{equation*}
    T(x,x') \ = \ H_a(x,x') (\Lambda_1(x) -
    \Lambda_1(x'))\e^{\delta |x_1'|} \; .
  \end{equation*}
  They vanish if $|x_1 - x_1'| \le |x_1'|$ since $x_1' \ge 0$
  (resp. $x_1' < 0$) then implies the same for $x_1$. Therefore
  \begin{equation*}
      \abs{T(x,x')} \ \le \  2 \abs{H_a(x,x')} \e^{\delta |x_1
      -x_1'|} \ \le \ 2 \abs{H_a(x,x')} \e^{\delta |x-x'|} \; ,
  \end{equation*}
  which together with $T(x,x)=0$ yields $|T(x,x')| \le C
  |H_a(x,x')| (\e^{\delta |x-x'|} - 1)$. Now \eqref{eq:45}
  follows from \eqref{eq:1} and \eqref{eq:43}. The
  estimate for $$\e^{\delta |x_2|} \Lambda_{2;a}(t) = \e^{\delta |x_2|}
  \e^{\im H_a t} \e^{-\delta |x_2|} \cdot \e^{\delta |x_2|} \Lambda_2
  \e^{-\im H_a t}$$
  follows from \eqref{eq:42} and from
  $\|\e^{\delta |x_2|} \Lambda_2 \| = \e^{\delta a}<\infty$.

  The proof of $\eqref{eq:41}$ is similar: Using
  \eqref{eq:20a} we write
  \begin{equation}\label{eq:47}
    \com{\rho(H_a)}{\Lambda_1} \ = \ \frac{1}{2 \pi} \int \di m(z)
    \partial_{\bar z} \rho(z) \com{R_a(z)}{\Lambda_1}
  \end{equation}
  and claim that
  \begin{equation}\label{eq:48}
    \norm{\com{R_a(z)}{\Lambda_1}\e^{\delta |x_1|} } \ \le \
    \frac{C}{|\Im z|^2}
  \end{equation}
  for $\delta=\delta(z)$ as in \eqref{eq:39}. Together with
  (\ref{eq:37}, \ref{eq:44}, \ref{eq:46}) this
  implies \eqref{eq:41}. To derive \eqref{eq:48}, note that the operator
  to be bounded is $-R_a(z)
  [H_a \, , \ \Lambda_1 ] \e^{\delta |x_1|} \cdot \e^{-\delta |x_1|} R_a(z)
  \e^{\delta |x_1|}$ and the bound follows from (\ref{eq:38}, \ref{eq:45}).

  The conclusion about $Z(a,t)$ follows from \eqref{eq:41} at $t=0$ and
  (\ref{eq:37}, \ref{eq:40}).\qed
\end{proof}

\subsection{Proof of Lemma \ref{lem:tracelimit}} It follows from
\eqref{eq:41} that $\Sigma_a'(t)$ is trace class. While \eqref{eq:48} holds
uniformly in $a$, including the Bulk case, \eqref{eq:46} fails in this
respect. Nevertheless $\Sigma_B'(t) \in \I_1$, since
\begin{equation}\label{eq:49}
    \sup_{a,B} \norm{\e^{\delta |x_2|} (\Lambda_{2;a}(t) - \Lambda_2)}
    \ \le \ C
\end{equation}
for $t$ in a compact interval.  In fact
\begin{equation*}
    \e^{\delta |x_2|} \left ( \Lambda_{2;a}(t) - \Lambda_2 \right )
    \ = \ \e^{\delta |x_2|} \int_0^t \e^{\im H_a s} \im
    \com{H_a}{\Lambda_2} \e^{-\im H_a s}
\end{equation*}
with
\begin{equation}\label{eq:50}
\sup_{a,B} \norm{\e^{\delta |x_2|} \e^{\im H_a t}
\com{H_a}{\Lambda_2} \e^{-\im H_a t} } \ \le \ C \; ,
\end{equation}
because of \eqref{eq:42} and of $\norm{\e^{\delta |x_2|}
\com{H_a}{\Lambda_2}} \le C$, c.f. \eqref{eq:45}.

To prove \eqref{eq:25} we use \eqref{eq:47} and $\J_a^* \J_a =1$ to write
\begin{multline*}
\J_a \Sigma_a'(t) \J_a^* \ = \ - \frac{1}{2\pi} \int \di m(z)
    \partial_{\bar z} \rho(z) \\
    \times \J_a \com{R_a(z)}{\Lambda_1}
    \e^{\delta |x_1|} \J_a^*  \cdot
    \e^{-\delta |x_1|} \e^{-\delta |x_2|} \cdot
    \J_a \e^{\delta |x_2|} \left ( \Lambda_{2;a}(t) - \Lambda_2
    \right ) \J_a^*
\end{multline*}
It is enough to establish convergence to the bulk expression pointwise in $z$,
since domination is provided by (\ref{eq:48}, \ref{eq:44}, \ref{eq:49},
\ref{eq:37}). We thus may show
\begin{align}\label{eq:51}
    \J_a \com{R_a(z)}{\Lambda_1} \e^{\delta |x_1|} \J_a^*
    \ &\xrightarrow[a \rightarrow \infty ]{s} \ \com{R_B(z)}{\Lambda_1}
    \e^{\delta |x_1|}\; , \\ \label{eq:52}
    \J_a  \left ( \Lambda_{2;a}(t) - \Lambda_2
    \right ) \J_a^* \e^{\delta |x_2|} \ &\xrightarrow[a \rightarrow \infty ]{s} \
     \left ( \Lambda_{2B}(t) - \Lambda_2
    \right ) \e^{\delta |x_2|} \; .
\end{align}
Since the l.h.s.'s are uniformly bounded in $a$ by (\ref{eq:48}, \ref{eq:49})
it suffices to prove convergence on the dense subspace of compactly supported
states in $\ell^2(\Z \times \Z)$, which amounts to dropping $\e^{\delta |x_i|}$
in (\ref{eq:51}, \ref{eq:52}). Eq. \eqref{eq:4a} implies the geometric
resolvent identity $\J_a R_a(z) - R_B(z) \J_a =  - R_B(z) E_a R_a(z)$, and by
taking the adjoint
\begin{equation*}
     \J_a R_a(z) \J_a^* - R_B(z) \ = \ - \left ( \J_a R_a(z) E_a^* + 1- \J_a
     \J_a^* \right ) R_B(z) \ \xrightarrow[a \rightarrow \infty ]{s}
     \ 0
\end{equation*}
because $E_a^* \xrightarrow[a \rightarrow \infty ]{s} 0$ by \eqref{eq:5} and
because $1- \J_a \J_a^* \xrightarrow[a \rightarrow \infty ]{s} 0$ is the
projection onto states supported in $\set{x_2 < -a}$.  This implies \cite[Thm.
VIII.20]{RS80}
\begin{equation}\label{eq:53}
    \operatorname*{s-lim}_{a \rightarrow \infty} \J_a f(H_a) \J_a^* \
    = \ f(H_B)
\end{equation}
for any bounded continuous function $f$, and in particular the modified limits
(\ref{eq:51}, \ref{eq:52}). The proof of \eqref{eq:26} is similar. We write the
integrand of $\J_a \Sigma_a''(t) \J_a^*$ as
\begin{equation*}
    \J_a \com{R_a(z)}{\Lambda_1} \e^{\delta |x_1|} \J_a^* \cdot
    \e^{-\delta |x_1|}\e^{-\delta |x_2|} \cdot \J_a \e^{\delta
    |x_2|} \com{H_a}{\Lambda_{2;a}(t)} R_a(z) \J_a^* \; .
\end{equation*}
Since the estimates for the first two factors have already been given, all we
need are
\begin{gather*}
  \sup_{a,B} \norm{\e^{\delta |x_2|} \com{H_a}{\Lambda_{2;a}(t)}} \
  \le \ C \; , \\
  \J_a \com{H_a}{\Lambda_{2;a}(t)} \J_a^* \ \xrightarrow[a \rightarrow
  \infty]{s} \ \com{H_B}{\Lambda_{2B}(t)} \; .
\end{gather*}
The first estimate is just \eqref{eq:50} and the second is again implied by
\eqref{eq:53}. \qed

\subsection{Proof of Lemma \ref{lem:sigmaBrep}} Let
$P_{\lambda_0}^\perp = 1 - P_{\lambda_0}$. By the definition \eqref{eq:4} we
have
\begin{equation*}
    \sigma_B(\lambda_0) \ = \ \im \tr \left ( P_{\lambda_0}
    \Lambda_1 P_{\lambda_0}^\perp \Lambda_2 P_{\lambda_0} -
    P_{\lambda_0} \Lambda_2 P_{\lambda_0}^\perp \Lambda_1
    P_{\lambda_0} \right ) \; .
\end{equation*}
Since the two terms are separately trace class by \eqref{eq:a}, we also have
$-\im \sigma_B(\lambda_0) = \tr T_{\lambda_0}$ with $T_{\lambda_0}$ as in
\eqref{eq:29a}; see \eqref{eq:44a}. Now \eqref{eq:30} yields
\begin{equation*}
    -\im \sigma_B(\lambda_0) \ = \ \tr \left ( E_- T_{\lambda_0} E_- \ + \
    E_+ T_{\lambda_0} E_+ \ + \ \sum_{\lambda \in \mathcal
    E_\Delta} E_{\{\lambda\}} T_{\lambda_0} E_{\{\lambda\}} \right ) \; ,
\end{equation*}
and the claim follows from
\begin{equation*}
    \tr P T_{\lambda_0} P = \tr P \com{P_{\lambda_0}}{\Lambda_1} \Lambda_2 P
\end{equation*}
for $P=P^*$ with $P P_{\lambda_0}^\perp =0$ or $P P_{\lambda_0} = 0 $, since
one or the other holds true for $P=E_{\pm}, E_{\{ \lambda\}}$.  Indeed, in the
first case, which also entails $P_{\lambda_0}^\perp P =0$, we have
\begin{equation*}
    \begin{split}
      P T_{\lambda_0} P \ = \ P P_{\lambda_0}
    \Lambda_1 P_{\lambda_0}^\perp \Lambda_2 P_{\lambda_0} P \ &= \ P \left ( P_{\lambda_0} \Lambda_1 \Lambda_2 - \Lambda_1
    P_{\lambda_0} \Lambda_2\right ) P  \\
    & = \ P \com{P_{\lambda_0}}{\Lambda_1} \Lambda_2 P \; .
    \end{split}
\end{equation*}
The other case is similar:
\begin{equation*}
    \begin{split}
      P T_{\lambda_0} P \ = \
      - P P_{\lambda_0}^\perp\Lambda_1 P_{\lambda_0} \Lambda_2 P_{\lambda_0}^\perp P \
      & = \ - P \left ( P_{\lambda_0}^\perp \Lambda_1 \Lambda_2 - \Lambda_1
    P_{\lambda_0}^\perp \Lambda_2 \right )  P \\
    &= \ - P \com{P_{\lambda_0}^\perp}{\Lambda_1} \Lambda_2 P \ = \
    P \com{P_{\lambda_0}}{\Lambda_1} \Lambda_2 P  \; . \quad \qed
    \end{split}
\end{equation*}

\subsection{Consequences of localization}We now discuss the technical
consequences of assumption \eqref{eq:2}. In fact, all that we say in this
section is a consequence of the following (weaker) estimate
\begin{equation}\label{eq:2epsilon} \sup_{g \in B_1(\Delta)} \sum_{x,x' \in
\Z^2}
    \abs{g(H_B)(x,x')} \e^{-\eps |x|} \e^{\mu |x-x'|} \ =: \ D_\eps \
    < \ \infty \; ,
\end{equation}
for every $\eps > 0$, where the factor $(1+|x|)^{-\nu}$  of \eqref{eq:2} has
been replaced by an exponential. Note that \eqref{eq:2epsilon} follows from
\eqref{eq:2} since $\e^{-\eps |x|} \le C_{\eps,\nu} (1+|x|)^{-\nu}$. (We
require \eqref{eq:2} to prove integrality of $2 \pi \sigma_B$ (Prop.\
\ref{prop:quantized} below), otherwise \eqref{eq:2epsilon}  would suffice for
the results described here.)

In terms of operators, rather than of matrix elements, \eqref{eq:2epsilon}
implies that for some $\mu
>0$ and all $\eps
>0$
\begin{equation}\label{eq:34}
    \sup_{g,\ell} \norm{\e^{\mu \ell(x)} \e^{-\eps |x|} g(H_B)
    \e^{-\mu \ell(x)}} \ \le \ D_\eps \ < \ \infty \; ,
\end{equation}
where the supremum with $g \in B_1(\Delta)$ is also taken over Lipschitz
functions $\ell$ as in \eqref{eq:39a}. In fact, the norm in \eqref{eq:34} is
estimated by Holmgren's bound \eqref{eq:43} as the larger of
\begin{equation}\label{eq:34a}
    \sup_{x} \sum_{x'} \e^{\mu (\ell(x) - \ell(x'))} \e^{-\eps
    |x|} \abs{g(H_B)(x,x')}
\end{equation}
and a similar quantity with $x,x'$ under the supremum and summation
interchanged.  After bounding the supremum by a sum, both quantities are
estimated by \eqref{eq:2epsilon}. Conversely, we take $\ell(x) = |x-x'|$ and
consider the $(x,x')$ matrix element of the operator in \eqref{eq:34},
\begin{equation}\label{eq:34b}
    \e^{\mu |x-x'|} \e^{- \eps |x|} \abs{g(H_B)(x,x')} \ \le \ D_\eps
    \; .
\end{equation}
The sum in \eqref{eq:2epsilon} is finite if $\mu$ is replaced there by $\mu/2$
and $\eps$ by $2 \eps$.

We say that a bounded operator $X$ is {\em confined in direction $i$} ($i=1,2$)
if for some $\delta > 0$ and all (small) $\eps > 0$
\begin{equation}\label{eq:new1}
    \norm{X}_{\eps, \delta}^{(i)} \ := \ \norm{X \e^{-\eps |x|} \e^{\delta |x_i|}}  \
    < \ \infty \; .
\end{equation}
Bounds of a similar form are (\ref{eq:48}, \ref{eq:49}), where a weight was
applied to an operator $X$, which could have as well been replaced by $X^*$.
Equivalently, the weight could have been placed on either side of $X$.  Here,
by contrast, dynamical localization will allow to establish \eqref{eq:new1} for
some operators $X$, but not for their adjoints. The asymmetry originates from
the following: if $X$ is confined, so are $BX$ for $B$ bounded and $X g(H_B)$
for $g \in B_1(\Delta)$, with
\begin{gather}\label{eq:new2}
    \norm{BX}_{\eps,\delta}^{(i)} \ \le \ \norm{B} \, \norm{X}_{\eps, \delta}^{(i)} \; , \\
    \label{eq:new3}
    \norm{Xg(H_B)}_{\eps,\delta}^{(i)}  \le \ D_{\frac{\eps}{2}}
    \norm{X}_{\frac{\eps}{2},\delta}^{(i)} \;
\end{gather}
for small $\delta > 0$.  In fact,
\begin{multline*}
    \norm{X g(H_B) \e^{-\eps |x|} \e^{\delta |x_2|} } \\ \le \ \norm{X
    \e^{-\frac{\eps}{2} |x|} \e^{\delta |x_2|}} \cdot \norm{ \e^{-(\delta |x_2|
    - \frac{\eps}{2} |x|)}  g(H_B) \e^{-\frac{\eps}{2} |x|} \e^{(\delta |x_2|
    - \frac{\eps}{2} |x|)}}  \; ,
\end{multline*}
and for sufficiently small $\eps, \delta  >0$ the Lipschitz norm of $\delta
|x_2| - \frac{\eps}{2} |x|$ is smaller than $\mu$, whence \eqref{eq:34}
applies.

\begin{lemma}\label{lem:confinement}
  Let $S \subset \R$ be a Borel set that either contains or is disjoint from
  $\set{\lambda | \lambda < \Delta}$ and similarly for $\set{\lambda | \lambda
  > \Delta}$, i.e., $E_S \in B_1(\Delta)$. Let $X$ be a confined operator in
  direction $i$ ($i=1,2$).
  \begin{itemize}
    \item[i)] The following operators are also confined in direction $i$,
    as indicated by the estimates
    \begin{gather} \label{eq:new5}
      \norm{\com{X}{g(H_B)}}_{\eps,\delta}^{(i)} \ \le \ C
      \norm{X}_{\frac{\eps}{2},\delta}^{(i)} \; , \quad (g \in B_1(\Delta)) \; ,
      \\ \label{eq:new4}
      \norm{E_S^\perp X E_S}_{\eps, \delta}^{(i)} \ \le \ C
      \norm{X}_{\frac{\eps}{2},\delta}^{(i)}   \; .
    \end{gather}
    \item[ii)] If in addition $S \subset \Delta$, then the following operators
    are also confined
    \begin{gather}\label{eq:new6}
      \norm{\com{H_B}{A_{T,B}(X)} E_S}_{\eps,\delta}^{(i)} \ \le \
      \frac{C}{T} \norm{X}_{\frac{\eps}{2},\delta}^{(i)} \; , \\ \label{eq:new7}
      \norm{\left ( A_{T,B}(X) - X \right ) E_S}_{\eps,\delta}^{(i)} \ \le \ C
      \norm{X}_{\frac{\eps}{2}, \delta}^{(i)} \; ,
    \end{gather}
    and given $S' \subset \R$ with $d = \dist(S,S') > 0$,
    \begin{equation}\label{eq:new8}
    \norm{E_{S'} A_{T,B}(X) E_S}_{\eps,\delta}^{(i)}
    \ \le \ \frac{C}{T} \norm{X}_{\frac{\eps}{2}, \delta}^{(i)} \;
    .
    \end{equation}
    \item[iii)] Properties (i, ii) also hold for $X = \Lambda_i$,
    with $\norm{X}_{\frac{\eps}{2},\delta}^{(i)}$ replaced by $1$.
  \end{itemize}
  The constants $C$ depend on $\eps,\delta$, but not on the remaining
  quantities, except for \eqref{eq:new8} which depends on $d$.
\end{lemma}

The main use of confined operators will be through the following remark: If
$X_i$, $(i=1,2)$, is confined in direction $i$, then $X_2 X_1^* \in \I_1$ with
\begin{equation}\label{eq:new9}
    \norm{X_2 X_1^*}_1 \ \le \ C \norm{X_2}_{\eps,\delta}^{(2)}
    \norm{X_1}_{\eps,\delta}^{(1)}
\end{equation}
for $2 \eps < \delta$.  In particular, if also $X_1^* X_2 \in \I_1$,
\eqref{eq:new9} is a bound for $\tr X_1^* X_2 = \tr X_2 X_1^*$.  Indeed,
\eqref{eq:new9} follows from $\e^{-\delta |x_2|}\e^{2 \eps |x|} \e^{-\delta
|x_1|} = \e^{-(\delta - 2 \eps)|x|} \in \I_1$.

\subsection{Proof of Lemma \ref{lem:confinement}} For $X$ confined,
(\ref{eq:new5}) is implied by (\ref{eq:new2}, \ref{eq:new3}). We thus consider
$X=\Lambda_i$, where it is enough to estimate
\begin{multline*}
    \com{\Lambda_i}{g(H_B)} \e^{- \eps |x|} \e^{\pm \delta x_i}
    \ = \ \Lambda_i g(H_B) (1 - \Lambda_i) \e^{-\eps |x|} \e^{\pm \delta x_i}
    \\ + (1 - \Lambda_i) g(H_B) \Lambda_i \e^{-\eps |x|} \e^{\pm  \delta x_i}
    \; .
\end{multline*}
In the $+$ case, for instance, the second term is bounded because $\Lambda_i
\e^{\delta x_i}$ is.  By \eqref{eq:34} this holds for the first one too.

{From} now on the switch functions and the confined operators will be treated
simultaneously. Eq.~\eqref{eq:new4} follows from  \eqref{eq:new5} and
$E_S^\perp X E_S = E_S^\perp \com{X}{E_S}$. To prove \eqref{eq:new6} we
consider
\begin{multline}\label{eq:new10}
    T \cdot \im \com{H_B}{A_{T,B}(X)} E_S \ = \ (\e^{\im H_B T} X \e^{-\im
    H_B T} - X ) E_S \\
    = \ \e^{\im H_B T} \left ( X \e^{-\im H_B T} E_S - \e^{-\im H_B T} E_S X
    \right ) E_S \ - \ E_S^\perp X E_S \; .
\end{multline}
The term in parentheses is bounded by \eqref{eq:new5} for $g(\lambda) =
\e^{-\im \lambda T} E_S(\lambda)$. The norm \eqref{eq:new1} of \eqref{eq:new10}
is uniformly bounded in $T \in \R$ by (\ref{eq:new2}, \ref{eq:new3},
\ref{eq:new4}).  The same bound applies to
\begin{equation*}
    ( A_{T,B}(X) - X) E_S \ = \ \frac{1}{T} \int_0^T \di t ( \e^{\im H_B t} X
    \e^{-\im H_Bt} - X ) E_S \; .
\end{equation*}

We now turn to \eqref{eq:new8}, which is related to an integration by parts
lemma of \cite{Ka58}. Since $S \subset \Delta$ and $d > 0$, there is a contour
$\gamma$ in the complex plane (of length $\le 4 |\Delta| + 2 d$) encircling $S$
once, but not $S'$, at a distance $\ge d/2$ from both.  Then
\begin{equation*}
    \widetilde X \ = \ \frac{1}{2 \pi} \int_\gamma \di z R(z) E_{S'} X E_S R(z)
\end{equation*}
is convergent in the norm \eqref{eq:new1} because of (\ref{eq:new2},
\ref{eq:new3}, \ref{eq:new4}) (note that $(2/d) \cdot E_S(\lambda)
(z-\lambda)^{-1} \in B_1(\Delta)$). Its commutator with $H_B$ is
\begin{equation*}
    \begin{split}
      \im \com{H_B}{\widetilde X} \ &= \ - \frac{1}{2\pi \im} \int_\gamma \di z
      \com{H_B -z}{R(z) E_{S'} X E_S R(z)} \\
      &= \ - \frac{1}{2\pi \im} \int_\gamma \di z ( E_{S'} X E_S R(z) - R(z)
      E_{S'} X E_S ) \ = \ E_{S'} X E_S \; .
    \end{split}
\end{equation*}
Therefore, $E_{S'} A_{T,B}(X) E_S = A_{T,B}(E_{S'} X E_S) = E_{S'} \im
\com{H_B}{A_{T,B}(\widetilde X)} E_S$ and the claim follows from
\eqref{eq:new6}. \qed

\subsection{Proof of Lemma \ref{lem:track2}} We first prove \eqref{eq:31b} and
begin by recalling, see (\ref{eq:23}, \ref{eq:12}), that
\begin{multline}\label{eq:new11}
    \frac{1}{T} \int_0^T \tr E_\Delta \Sigma_B''(t) E_\Delta \ =  \
    \frac{\im}{2\pi} \int \di m(z) \partial_{\bar z} \rho(z) \tr E_\Delta
    R(z) \com{H_B}{\Lambda_1} \cdot \\
    \cdot R(z) \com{H_B}{A_{T,B}(\Lambda_2)} R(z) E_\Delta \; .
\end{multline}
By (\ref{eq:new2}, \ref{eq:new3}, \ref{eq:new6}) we have for small $\delta >0$
\begin{equation*}
    \norm{R(z) \com{H_B}{A_{T,B}(\Lambda_2)} R(z) E_\Delta}_{\eps,\delta}^{(2)}
    \ \le \ \frac{C}{T}
    |\Im z|^{-2} \; ,
\end{equation*}
and, together with \eqref{eq:45},
\begin{equation*}
    \norm{ \com{H_B}{\Lambda_1} R(z)E_\Delta}_{\eps,\delta}^{(1)}
    \ \le \ C |\Im z|^{-1} \; .
\end{equation*}

By \eqref{eq:new9} the trace in \eqref{eq:new11} is bounded by a constant times
$T^{-1} |\Im z|^{-3}$.  As the constant is independent of $z$, \eqref{eq:31b}
now follows by means of \eqref{eq:37}.

The operator under the trace in \eqref{eq:31a} is
\begin{multline}\label{eq:new12}
    E_\Delta P_{\lambda_0} \Lambda_1 ( A_{T,B}(\Lambda_2) - \Lambda_2 )
    E_\Delta - E_\Delta \Lambda_1 P_{\lambda_0} (A_{T,B}(\Lambda_2) - \Lambda_2
    ) E_\Delta \\
    = E_\Delta P_{\lambda_0} \Lambda_1 P_{\lambda_0}^\perp \cdot
     ( A_{T,B}(\Lambda_2) - \Lambda_2 )
    E_\Delta \\ - E_\Delta P_{\lambda_0}^\perp   \Lambda_1 P_{\lambda_0}\cdot
     ( A_{T,B}(\Lambda_2) - \Lambda_2 )
    E_\Delta \; .
\end{multline}
We claim that the two terms on the r.h.s.\ are separately trace class. In fact
\eqref{eq:new4} implies $
    \| P_{\lambda_0} \Lambda_1 P_{\lambda_0}^\perp \e^{-\eps |x|} \e^{\delta
    |x_1|}\|  \le  C $,
and similarly with $P_{\lambda_0}, P_{\lambda_0}^\perp$ interchanged, and the
bound \eqref{eq:49} also applies with $A_{T,B}(\Lambda_2)$ in place of
$\Lambda_{2,B}(t)$. (Note however that the bound so obtained is not uniform in
$T$.)

A factor $P_{\lambda_0}$, resp.\ $P_{\lambda_0}^\perp$, may now be cycled
around the traces of the two terms on the r.h.s.\ of \eqref{eq:new12}. The
trace \eqref{eq:31a} thus equals \begin{multline}\label{eq:new13}
    \tr E_\Delta P_{\lambda_0} \Lambda_1 P_{\lambda_0}^\perp \cdot P_{\lambda_0}^\perp
    A_{T,B}(\Lambda_2)
    P_{\lambda_0} E_\Delta \\
     - \tr E_\Delta P_{\lambda_0}^\perp   \Lambda_1 P_{\lambda_0}\cdot
     P_{\lambda_0} A_{T,B}(\Lambda_2)  P_{\lambda_0}^\perp
     E_\Delta - \tr E_\Delta T_{\lambda_0} E_\Delta \; ,
\end{multline}
where we used that the two terms of $T_{\lambda_0}$, see \eqref{eq:29a}, are
separately trace class.

We next show that the first two terms of \eqref{eq:new13} are uniformly bounded
in $\lambda_0 \in \Delta$, $T > 0$. Indeed, $X_1 = P_{\lambda_0}^\perp
\Lambda_1 P_{\lambda_0} E_\Delta$ and $X_2 = P_{\lambda_0}^\perp
     A_{T,B}(\Lambda_2)
    P_{\lambda_0} E_\Delta = P_{\lambda_0}^\perp
    ( A_{T,B}(\Lambda_2) - \Lambda_2 )
    P_{\lambda_0} E_\Delta + P_{\lambda_0}^\perp \Lambda_2
P_{\lambda_0} E_\Delta$ are uniformly confined by (\ref{eq:new4},
\ref{eq:new7}) and the conclusion is by \eqref{eq:new9}.

Finally, we will show that these two terms vanish as $T \rightarrow \infty$,
pointwise in $\lambda_0 \in \Delta$.  The first one is split according to
$P_{\lambda_0} = P_\lambda + (P_{\lambda_0} - P_\lambda)$ for any $\lambda <
\lambda_0$, $\lambda \in \Delta$:
\begin{multline*}
  \tr P_{\lambda_0}^\perp A_{T,B}(\Lambda_2) P_{\lambda_0} E_\Delta \cdot
  E_\Delta P_{\lambda_0} \Lambda_1 P_{\lambda_0}^\perp \\
  \begin{aligned}
    & = \ \tr P_{\lambda_0}^\perp A_{T,B}(\Lambda_2)
    P_{\lambda}  E_\Delta \cdot
  E_\Delta P_{\lambda_0} \Lambda_1 P_{\lambda_0}^\perp \\ & \quad + \tr
   P_{\lambda_0}^\perp A_{T,B}(\Lambda_2) P_{\lambda_0} E_\Delta \cdot
    (P_{\lambda_0} -  P_{\lambda} ) \cdot
  E_\Delta P_{\lambda_0}\Lambda_1 P_{\lambda_0}^\perp \\
  & \equiv \mathrm{I} \ + \ \mathrm{II} \; .
  \end{aligned}
\end{multline*}

In $\mathrm{II}$, we extract the weights of the confined operators, so that the
middle factor becomes
\begin{multline*}
    \e^{2 \eps |x|} \e^{-\frac{\delta}{2} ( |x_1| + |x_2| )} \cdot \e^{-\eps
    |x|} e^{\frac{\delta}{2} ( |x_1| - |x_2| )} (P_{\lambda_0} -  P_{\lambda} )
    \e^{\frac{\delta}{2} ( |x_2| - |x_1| )} \e^{-\eps
    |x|} \cdot
    \\ \cdot \e^{-\frac{\delta}{2} ( |x_1| + |x_2| )} \e^{2 \eps |x|} \; .
\end{multline*}
For $\delta/2 > 2 \eps$ the operators on the sides are trace class, and the
middle one is uniformly bounded in $\lambda \in \Delta$ by \eqref{eq:34}.
Moreover, it converges weakly to zero as $\lambda \uparrow \lambda_0$, as this
holds true by $P_{\lambda_0}- P_\lambda \xrightarrow[]{s} 0$ for matrix
elements between states from the dense subspace of compactly supported states
in $\ell^2(\Z^2)$. Using
\begin{equation*}
    X_n \xrightarrow[]{w} 0 \; , \quad Y_1 \, , \ Y_2 \in \I_1 \quad \Longrightarrow
    \quad
    \norm{Y_1 X_n Y_2}_1 \rightarrow 0 \; ,
\end{equation*}
we conclude that $\mathrm{II}$ can be made uniformly small in $T$ by picking
$\lambda$ close to $\lambda_0$. The term $\mathrm{I}$ is then seen to be
$\mathcal O(T^{-1})$ by \eqref{eq:new8} with $S = (-\infty, \lambda) \cap
\Delta$ and $S' = [\lambda_0,\infty)$.

The second trace in \eqref{eq:new13} is dealt with slightly differently. We
insert $P_{\lambda_0} = P_\lambda + E_\Delta (P_{\lambda_0} - P_\lambda )
E_\Delta$ for $\lambda < \lambda_0$, $\lambda \in \Delta$, which yields two
well-defined traces. The second can be made uniformly small in $T$, as was the
case for $\mathrm{II}$ above. The first one, which by \eqref{eq:44a} equals
$\tr P_\lambda A_{T,B}(\Lambda_2) P_{\lambda_0}^\perp E_{\Delta} \cdot
E_{\Delta} P_{\lambda_0}^\perp \Lambda_1 P_\lambda$, is $\mathcal O(T^{-1})$ by
\eqref{eq:new8}, this time with $S=[\lambda_0,\infty)\cap \Delta$, $S' =
(-\infty, \lambda)$. \qed

\subsection{Proof of Lemma \ref{lem:uniformintbound}}
We shall need a particular choice of basis $\set{\psi_{\lambda;j}}$ for $\ran
E_{\set{\lambda}}$, which is related to a SULE basis \cite{dRJLS96}. (The issue
is only of relevance if $\lambda \in \mathcal E_\Delta$ is degenerate, since
otherwise $\psi_\lambda$ is unique up to a phase.) We claim a basis can be
chosen so that \eqref{eq:34} applies not only to $g(H_\lambda) =
E_{\set{\lambda}} = \sum \psi_{\lambda;j} \left ( \psi_{\lambda;j}\ , \ \cdot \
\right )$, but also to the rank one projections into which it is decomposed
(upon changing $\mu, D_\eps$, depending on $C_4$). Since $\norm{\phi \left (
\psi \, \ \cdot \ \right ) } = \norm{\phi} \norm{\psi}$, this amounts to
\begin{equation}\label{eq:34c}
    \sup_{\ell} \norm{\e^{\mu \ell(x)} \e^{-\eps |x|} \psi_{\lambda;j}} \norm{
    \e^{-\mu \ell(x)} \psi_{\lambda;j}} \ \le \ D_\eps \; .
\end{equation}

In fact, since $\sum_{x} E_{\set{\lambda}}(x,x) \ = \ \tr E_{\set{\lambda}} \
\le \ C_4$, we may pick $x_0 \in \Z^2$ such that $E_{\set{\lambda}}(x_0,x_0) =
\max_{x} E_{\set{\lambda}} (x,x)$. Let $\psi_{\lambda;0}(x) =
E_{\set{\lambda}}(x,x_0)/ E_{\set{\lambda}}(x_0,x_0)^{1/2}$. This normalized
eigenfunction satisfies the bounds
\begin{equation*}
    \abs{\psi_{\lambda;0}(x)} \ \le \ \begin{cases}
      D_\eps \e^{\eps |x_0|} \e^{-\mu |x - x_0|} /
      E_{\set{\lambda}}(x_0,x_0)^{1/2} \; , \\
      E_{\set{\lambda}}(x_0,x_0)^{1/2} \; .
    \end{cases}
\end{equation*}
The first one follows from \eqref{eq:34b} for $g(H_B)= E_{\set{\lambda}}$, and
the second from $$\abs{E_{\set{\lambda}}(x,x_0)} \le
E_{\set{\lambda}}(x,x)^{1/2} E_{\set{\lambda}}(x_0,x_0)^{1/2} \le
E_{\set{\lambda}}(x_0,x_0) \; .$$ Combining them into a geometric mean yields
$\abs{\psi(x)} \le D_\eps^{\half} \e^{\frac{\eps}{2} |x_0|} \e^{- \frac{\mu}{2}
|x-x_0|}$ and, by the triangle inequality,
\begin{equation*}
      \abs{ \psi_{\lambda;0}(x) \overline \psi_{\lambda;0}(x')} \ \le  \
      D_\eps
      \e^{\eps |x_0|} \e^{-\frac{\mu}{2}\left ( |x-x_0|+ |x'-x_0| \right )}\
        \le \ D_\eps \e^{\eps |x|} \e^{- \left ( \frac{\mu}{2} - \eps \right )
        |x-x'|} \; .
\end{equation*}
For small $\eps$ the bound \eqref{eq:34b} is reproduced for $\psi_{\lambda;0}
\left ( \psi_{\lambda;0} \ , \ \cdot \ \right )$ in place of
$E_{\set{\lambda}}$, with a smaller value of $\mu$. Since the rank of
$E_{\set{\lambda}} - \psi_{\lambda;0} \left ( \psi_{\lambda;0} \ , \ \cdot \
\right )$ is one less than the rank of $E_{\set{\lambda}}$, the task is
completed by induction.

After these preliminaries, we turn to the proof of Lemma
\ref{lem:uniformintbound} proper. We denote by $\widetilde { \mathcal
E}_\Delta$ the eigenvalues in $\mathcal E_\Delta$ listed according to
multiplicity. More precisely, we let $\widetilde {\mathcal E}_\Delta$ be the
set of pairs $\zeta = (\lambda ; n)$ with $\lambda \in \mathcal E_\Delta$ and
$n$ a non-negative integer less than the multiplicity of $\lambda$. The
eigenvectors $\{\psi_\zeta$, $\zeta \in \widetilde { \mathcal E}_\Delta \}$
constructed above are an ortho-normal basis for $\mathrm{ran} E_{\Delta}$.

Let, for $\zeta \in \widetilde { \mathcal E}_\Delta$,
\begin{equation*}
    M_\zeta \ = \ \min \left ( \norm{\Lambda_1 \psi_\zeta},
    \norm{(1-\Lambda_1) \psi_\zeta}, \norm{\Lambda_2
    \psi_\zeta}, \norm{(1-\Lambda_2) \psi_\zeta} \right ) \; .
\end{equation*}
We claim that
\begin{equation}\label{eq:35}
    \sum_{\zeta \in \widetilde { \mathcal E}_\Delta} M_\zeta \ < \ \infty \;
    .
\end{equation}
This states that almost all eigenfunctions are localized in at least one among
the left, right, upper, and lower half planes, and hence in at most two
(intersecting) ones.  In particular almost no eigenfunction encircles the
origin, which makes them insensitive to a flux tube applied there \textemdash a
fact used in some explanations \cite{Ha82,Pr87} of the QHE.

We apply \eqref{eq:34c} to $\psi_\zeta \left ( \psi_\zeta, \cdot \right )$ and
use that for rank one operators $\norm{\phi \left ( \psi, \cdot \right ) } =
\norm{\phi}\norm{\psi}$ to obtain $\| \e^{\mu \ell(x)} \e^{-\eps |x|}
\psi_\zeta \|  \|
    \e^{-\mu \ell(x)} \psi_\zeta\|  \le D_\eps$.
For $\ell(x) = x_1$ we have $\Lambda_1(x) \le \e^{-\mu \ell(x)}$,
implying
\begin{equation*}
    C_2 \norm{\Lambda_1 \psi_\zeta}^{-1} \ \ge \
    \norm{\e^{\mu x_1} \e^{-\eps |x| }\psi_\zeta} \; ,
\end{equation*}
similar estimates for $1-\Lambda_1$, $\Lambda_2$, and $1-\Lambda_2$ have $x_1$
on the r.h.s.\ replaced by $-x_1$, $x_2$, and $-x_2$ respectively. Therefore,
\begin{equation*}
    \begin{split}
      M_\zeta^{-2} \ &= \ \max \left ( \norm{\Lambda_1 \psi_\zeta}^{-2},
    \norm{(1-\Lambda_1) \psi_\zeta}^{-2}, \norm{\Lambda_2
    \psi_\zeta}^{-2}, \norm{(1-\Lambda_2) \psi_\zeta}^{-2} \right
    )\\
    &\ge \ \frac{1}{4} \left ( \norm{\Lambda_1 \psi_\zeta}^{-2} +
    \norm{(1-\Lambda_1) \psi_\zeta}^{-2} + \norm{\Lambda_2
    \psi_\zeta}^{-2} + \norm{(1-\Lambda_2) \psi_\zeta}^{-2} \right
    ) \\
    &\ge \ \frac{1}{4 C_2^2} \left ( \psi_\zeta, \e^{-2 \eps |x|}
    \left ( \e^{2\mu x_1} + \e^{-2\mu x_1} + \e^{2\mu x_2} + \e^{-2\mu
    x_2} \right ) \psi_\zeta \right ) \\
    &\ge \ \frac{1}{4 C_2^2} \left ( \psi_\zeta, \e^{(\mu-2 \eps) |x|} \psi_\zeta \right
    ),
    \end{split}
\end{equation*}
where we use $\e^{2 \mu |x_1|} + \e^{2 \mu |x_2|} \ge \e^{\mu ( |x_1| +
|x_2|)}$. Now let $\eps > 0$ be small enough that $\delta := \mu - 2 \eps > 0$.
Then
\begin{equation*}
    M_{\zeta} \ \le \ 2 D_{\varepsilon} \left [ \left ( \psi_{\zeta},
    \e^{\delta |x| } \psi_{\zeta} \right ) \right ]^{-
    \frac{1}{2}} \ \le \ 2 D_{\varepsilon} \left ( \psi_{\zeta},
    \e^{-\frac{1}{2}\delta |x| } \psi_{\zeta} \right ),
\end{equation*}
where in the last step we have applied Jensen's inequality with the convex
function $t \mapsto t^{-\frac{1}{2}}$. As $\{ \psi_{\zeta} : \zeta \in
\widetilde{\mathcal E}_{\Delta}\}$ are ortho-normal, we conclude that
\begin{equation*}
    \sum_{\zeta \in \widetilde{\mathcal E}_{\Delta}} M_{\zeta} \ \le
    \ 2 D_{\varepsilon} \tr \e^{-\frac{1}{2} \delta |x|} \ < \ \infty ,
\end{equation*}
proving (3.37).

We can now estimate the traces in \eqref{eq:32}:
\begin{equation}\label{eq:36}
    \abs{\tr E_{\set{\lambda}} \e^{-\im H_B t}
    \com{H_B}{\Lambda_1} \e^{\im H_Bt} \Lambda_2 E_{\set{\lambda}} } \ \le \
    \sum_{\zeta=(\lambda; \cdot)} \abs{\left ( \psi_\zeta,
    \com{H_B}{\Lambda_1} \e^{\im H_Bt} \Lambda_2 \psi_\zeta
    \right ) } \; .
\end{equation}
By inserting $\Lambda_2 = 1 - (1 -\Lambda_2)$, the terms on the right hand side
may also be expressed as
\begin{equation*}
    \abs{\left ( \psi_\zeta,
    \com{H_B}{\Lambda_1} \e^{\im H_Bt} (1- \Lambda_2) \psi_\zeta
    \right ) } \; .
\end{equation*}
Using
\begin{equation*}
   \left ( \psi_\zeta, \com{H_B}{\Lambda_1} \phi \right ) \ = \
    \ \left ( \Lambda_1 \psi_\zeta, (\lambda - H_B ) \phi \right
    ) \ = \ - \left ( (1-\Lambda_1) \psi_\zeta, (\lambda - H_B) \phi
    \right ) \; ,
\end{equation*}
one sees that \eqref{eq:36} is bounded by a constant times $ \sum_{\zeta =
(\lambda;\cdot)} M_\zeta$, so the right hand side of \eqref{eq:32} is bounded
by $\sum_\zeta M_\zeta$. \qed

\section{Analysis of the Harper Hamiltonian}\label{sec:harper}

In this section we prove Theorem \ref{thm:harper} which shows that the
contribution from bulk states in \eqref{eq:11b} can be non-zero. We begin with
the following proposition:
\begin{proposition}\label{prop:Cauchyaverages}
  Let $f(\set{V_x}_{x \in \Z^d})$ be a function which is bounded and continuous
  in the product topology on $\set{\set{V_x}_{x \in \Z^d} | \Im V_x \le 0} =
  {\overline \C_-}^{\Z^d}$. If $f$ is separately analytic in each $V_x$, then
  \begin{equation}\label{eq:Cauchyaverages}
    \Ev{f} \ = \ f(\set{-\im}_{x \in \Z^d}) \; ,
  \end{equation}
  where $\Ev{\cdot}$ represents the average with respect to the product measure
  $$\di {\mathbb P}(\set{V_x}_{x \in \Z^d}) \ := \ \prod_{x \in \Z^d} \frac{ \di V_x}{\pi (1 +V_x^2)} \; ,$$
  supported on $\set{\set{V_x}_{x \in \Z^d} | V_x \in \R} = \R^{\Z^d}$. The
  same statement holds for $\C_+$, $+ \im$  in place of $\C_-$, $-\im$.
\end{proposition}
\begin{proof}
  Let $S_j$ be an increasing sequence of finite sets with $\lim_j S_j = \cup_j S_j =
  \Z^d$, and let $\mathcal F_j^c$ denote the $\sigma$-algebra generated by
  $\set{V_x}_{x \in S_j^c}$.  So conditional expectation with respect to
  $\mathcal F_j^c$ is given by ``averaging out'' the variables
  $\set{V_x}_{x \in S_j}$. Thus
  \begin{equation*}
    f_j(\set{V_x}_{x \in S_j^c}) \ := \ \Ev{f|\mathcal F_j^c} \ = \
    \int \prod_{x \in S_j} \frac{ \di V_x}{\pi (1 +V_x^2)}
    f(\set{V_x}_{x \in S_j}\times\set{V_x}_{x \in S_j^c}) \; .
  \end{equation*}
  Because $f$ is bounded and separately analytic in each $V_x$, we may evaluate
  the integrals on the right hand side by residues to obtain $$
    f_j(\set{V_x}_{x \in S_j^c})  \ = \ f(\set{-\im}_{x
    \in S_j} \times \set{V_x}_{x \in S_j^c}) \; .$$
  Because $f$ is continuous and
   $\lim_{j \rightarrow \infty} \set{-\im}_{x
    \in S_j} \times \set{V_x}_{x \in S_j^c}   =   \set{-\im}_{x \in \Z^d}$
  in the product topology on ${\overline \C_-}^{\Z^d}$, we have
  \begin{equation*}
    \lim_{j \rightarrow \infty} f_j(\set{V_x}_{x \in S_j^c}) \ = \
    f(\set {-\im}_{x
    \in \Z^d})
  \end{equation*}
  for any $\set{V_x}_{x \in \Z^d} \in \R^{\Z^d}$. Since $f_j$ are
  uniformly bounded and $\Ev{f_j} =
  \Ev{f}$ for every $j$, we conclude by dominated convergence that
  \eqref{eq:Cauchyaverages} holds. \qed
\end{proof}

Turning now to the proof of Theorem \ref{thm:harper}, we first recall that, by
Lemma \ref{lem:tracelimit},
\begin{equation}\label{eq:theconvergence}
   -\frac{\im}{2}  \lim_{a\rightarrow \infty} \tr \rho'(H_a) \set{
   \com{H_a}{\Lambda_1} \,  , \, \Lambda_2 } \ = \ \Re \tr \Sigma''_B(0) \; ,
\end{equation}
where
\begin{equation*}
    \im \Sigma''_B(0) \ =  - \frac{1}{2 \pi} \int \di m (z) \partial_{\bar z}
    \rho(z) \tr \underbrace{R_B(z) \com{H_\phi}{\Lambda_1} R_B(z) \com{H_\phi}{\Lambda_2}
    R_B(z)}_{T_B(z)} \; .
\end{equation*}
In going from \eqref{eq:23} to the above expression for $\Sigma''_B(0)$ we have
replaced $H_B$ by $H_\phi$ in the commutators $\com{H_B}{\Lambda_i}$ since the
random potential commutes with each switch function $\Lambda_i$.

By Lemma \ref{lem:tracelimit}, we have $
    \sup_{a } \abs{  \tr \rho'(H_a)
    \set{\, \com{H_a}{\Lambda_1}, \Lambda_2 }} \ \le \ C \ < \ \infty$,
with a constant $C$ that depends on $\rho$ and on the bounds $C_1,C_3$ in
(\ref{eq:1}, \ref{eq:5}), but \underline{not} on the random constant $C_2$ in
\eqref{eq:2}. Since the constants $C_1,C_3$ are non-random in our setup, the
expectation in \eqref{eq:RadNik} is well defined, and furthermore can be
exchanged with the limit.

We claim that for $\Im z \neq 0$
\begin{equation}\label{eq:TCauchyaverage}
    \Ev{\tr T_B(z)} \ = \ \tr T_\phi(z + \im \alpha  \sigma(z))\; ,
\end{equation}
where $T_\phi(z) =  R_{\phi}(z) \com{H_{\phi}}{\Lambda_1} R_{\phi}(z)
    \com{H_{\phi}}{\Lambda_2} R_{\phi}(z) $,
with $R_\phi(z) \ = \ (H_\phi - z)^{-1}$, and $\sigma(z)= \Im z/|\Im z|$
denotes the sign of the imaginary part of $z$. Indeed, for $\Im z
>0$, it suffices to verify that $f_z(\set{V_x}) \ = \ \tr T_B(z)$ obeys the
hypotheses of Proposition \ref{prop:Cauchyaverages}. For that purpose, it is
useful to note that
$$ G_z(\set{V_x}_{x \in \Z^d}) \ := \ (H_\phi + \alpha V - z)^{-1}$$
is a continuous map from $\set{\set{V_x}_{x \in \Z^d} | \Im V_x \le 0}$ to the
bounded operators on $\ell^2(\Z^2)$ endowed with the strong operator topology.
Indeed, $z$ is in the resolvent set of $H_\phi + \alpha V$ since the numerical
range of this operator is contained in the closed lower half plane. Thus $G_z$
is well defined, SOT-continuous (since $\set{V_x}_x \mapsto H_\phi + \alpha V$
and $A \mapsto A^{-1}$ are SOT-continuous), and
\begin{equation}\label{eq:4.3a}\norm{G_z(\set{V_x}_{x \in \Z^d})} \ \le \
\frac{1}{\dist(z,\text{num. range}(H_\phi + \alpha V))} \ \le \ \frac{1}{|\Im
z|} \; .
\end{equation}

Furthermore, the Combes-Thomas bound \eqref{eq:38} extends to $G_z$, i.e.,
\begin{equation}\label{eq:extendedCT}
    \norm{\e^{\delta \ell(x) } G_z(\set{V_x}_{x \in \Z^d}) \e^{-\delta \ell(x)}} \
    \le \ \frac{C}{|\Im z|}\; , \quad \delta^{-1} = C \left ( 1 + |\Im z|^{-1}
    \right ) \; ,
\end{equation}
with $\ell(x)$  as in \eqref{eq:39a}. The resolvent of $\e^{\pm \delta \ell(x)}
(H_\phi  + \alpha V)\e^{\mp \delta \ell(x)}$, considered as a perturbation of
$H_\phi + \alpha V$, is in fact as stable as in \eqref{eq:38} where $H_\phi$
was self-adjoint, since the same bound \eqref{eq:4.3a} still holds for $\Im z
> 0$. Furthermore, we see in this way that
$$\set{V_x}_{x \in \Z^d} \mapsto \e^{\delta\ell(x) } G_z(\set{V_x}_{x \in
\Z^d}) \e^{-\delta \ell(x)}$$ is SOT-continuous.

Thus, for $\Im z > 0$,
\begin{multline*}
    \tr T_B(z) \ = \ \tr G_z(\set{V_x}_{x \in \Z^d})
    \com{H_\phi}{\Lambda_1}
    \e^{\delta |x_1|}  \ \cdot \ \e^{-\delta |x_1|}
    G_z(\set{V_x}_{x \in \Z^d})\e^{\delta |x_1|} \ \cdot  \\ \cdot \
    \e^{-\delta (|x_1| + |x_2|)}  \ \cdot \e^{\delta |x_2|}
    \com{H_\phi}{\Lambda_2} G_z(\set{V_x}_{x \in \Z^d}) \; ,
\end{multline*}
is a continuous function, which is bounded by
\begin{equation}\label{eq:Tbound}
\begin{split}
    \abs{\tr T_B(z)} \  & \le \
    \frac{C}{\delta^2} \ \norm{G_z(\set{V_x}_{x \in \Z^d})}^2 \,
    \norm{ \com{H_{\phi}}{\Lambda_1}
    \e^{\delta |x_1|}} \cdot \\
    & \qquad  \cdot \norm{\e^{-\delta |x_1|}G_z(\set{V_x}_{x \in \Z^d})
    \e^{\delta |x_1|}} \norm{\e^{\delta |x_2|}
    \com{H_{\phi}}{\Lambda_2}
    }  \\
     & \le \
    C \ \frac{(1+ |\Im z|^{-1})^2 }{ |\Im z|^3} \; ,
\end{split}
\end{equation}
with the factor of $1/\delta^2$ coming from the estimate \eqref{eq:44} on the
trace of $\e^{-\delta |x|}$. A similar argument is used for $\Im z < 0$. Since
the separate analyticity of $f_z(\cdot)=\tr T_B(z)$ is clear, Proposition
\ref{prop:Cauchyaverages} applies.

We see that
\begin{equation}\label{eq:CauchyaverageHelfSjo}
\Ev{\Re\tr\Sigma''_B(0)} \
    = \ - \frac{1}{2 \pi} \Im \int \di
    m(z) \partial_{\bar z} \rho(z) \tr T_\phi(z + \im \alpha \sigma(z) )
    \; ,
\end{equation}
where the interchange of $\int \di m(z)$ and $\E$ is justified by Fubini's
theorem and \eqref{eq:Tbound} since we may arrange for $\partial_{\bar z}
\rho(z)$ to vanish faster than $|\Im z|^5$ as $z$ approaches the real axis. We
note that
\begin{equation}\label{eq:4.6a}
    \abs{\tr T_\phi(z + \im \alpha \sigma(z))} \ \le \ \frac{C_\alpha}{[x^2 +
    (|y|+ \alpha)^2 ]^{3/2}} \; .
\end{equation}
In fact, now that $V=0$, $|\Im z|^{-1}$ in \eqref{eq:4.3a} may be replaced by
$\dist(z, \sigma(H_\phi))^{-1} \ \le \ \dist(z, [-2,2])^{-1}$ and the same
replacement carries over to the denominator in the estimate \eqref{eq:Tbound}
for $\tr T_\phi(z)$.

The only singularities in the integrand on the right hand side of
\eqref{eq:CauchyaverageHelfSjo} are jump discontinuities at $\Im z =0$.
Integrating by parts, on the upper and lower half planes separately, we find
\begin{equation}\label{eq:CauchyaverageHelfSjoa}
   \Ev{\Re\tr\Sigma''_B(0)} \ = \ \frac{1}{2 \pi} \Re \int_{-\infty}^\infty \di
    x \rho(x) \tr \left ( T_\phi(x + \alpha \im ) - T_\phi(x - \alpha \im)
    \right ) \; ,
\end{equation}
since by \eqref{eq:4.6a} there are no contributions from the boundary at
infinity. Upon writing $\rho(x) = -\int_x^\infty \rho'(\lambda) \di \lambda$,
and interchanging $\lambda$ and $x$ integration we obtain
\begin{equation}\label{eq:CauchyaverageHelfSjob}
   \Ev{\Re\tr\Sigma''_B(0)} \ = \ - \frac{1}{2 \pi}  \int_{-\infty}^\infty \di
    \lambda \rho'(\lambda) \int_{-\infty}^\lambda \Re \tr \left (
    T_\phi(x + \alpha \im ) - T_\phi(x - \alpha \im)
    \right ) \di x\; .
\end{equation}
This proves \eqref{eq:RadNik} with
\begin{equation}\label{eq:jBrepa}
    j_B(\lambda) \ = \ \frac{1}{2 \pi} \Re
    \int_{-\infty}^\lambda \tr \left ( T_\phi(x + \alpha \im ) -
    T_\phi(x - \alpha \im)
    \right ) \di x \; .
\end{equation}

To obtain the asymptotic expression \eqref{eq:asymptotic}, note that for
$|\lambda| > 2$
\begin{equation}\label{eq:jBrepb}
    j_B(\lambda) \ = \ \frac{1}{2 \pi} \Re  \int_{-\alpha}^\alpha \im \di \eta \tr
    T_\phi(\lambda + \im \eta)  \; ,
\end{equation}
because the difference of the right hand sides of (\ref{eq:jBrepa},
\ref{eq:jBrepb}) is the real part of an integral around a closed contour, which
may be deformed to infinity, of the analytic function $ \tr T_\phi(z)$, which
vanishes like $1/|z|^2$ as $z \rightarrow \infty$. (It is of interest to note
that for $\lambda$ in an internal gap of the spectrum of $H_\phi$,  the
corresponding contour integral gives the Bulk conductance
$\sigma_B^{(\phi)}(\lambda)$ for the Hamiltonian $H_\phi$ at Fermi energy
$\lambda$, so $ j_B(\lambda) = \sigma_{B}^{(\phi)}(\lambda) + \frac{1}{2 \pi}
\Re \im \int_{-\alpha}^\alpha \di \eta \tr T_\phi(\lambda + \im \eta)$.)

It is useful to rewrite \eqref{eq:jBrepb} as
\begin{equation}\label{eq:JBrepc}
    j_B(\lambda) \ = \  \frac{1}{2 \pi} \Re \int_{0}^\alpha \im \di \eta \left ( \tr
    T_\phi(\lambda + \im \eta) - \overline{\tr T_\phi(\lambda - \im \eta)} \right ) \; ,
\end{equation}
which follows by considering the contributions from $\eta <0$ and $\eta >0$
separately, and using $\Re \im \, w = - \Re \im \, \overline{w}$.

We obtain \eqref{eq:asymptotic} from the series for $T_B(\lambda + \im \eta) -
\overline{\tr T_B(\lambda - \im \eta)}$ produced by expanding each resolvent in
a Neumann series. For sufficiently large $|\lambda|$,
\begin{equation}\label{eq:neumann}
    R_\phi(\lambda + \im \eta) \ = \ - \frac{1}{\lambda}  \sum_{n=0}^\infty
        \left [ \frac{H_\phi - \im \eta}{\lambda} \right ]^n
\end{equation}
is  absolutely convergent, and
\begin{equation*}
\begin{aligned}
    \tr T_\phi(\lambda + \im \eta) \ = \ - \frac{1}{ \lambda^3}
    \sum_{N=0}^\infty \frac{1}{ \lambda^N} & \sum_{n_1 + n_2 + n_3 =
    N}  \tr  \left ( H_\phi - \im \eta \right )^{n_1}
    \com{H_\phi}{\Lambda_1} \cdot \\ & \quad  \quad \quad
    \cdot \left ( H_\phi - \im \eta \right )^{n_2}
    \com{H_\phi}{\Lambda_2}
    \left ( H_\phi - \im \eta \right )^{n_3} \; ,
\end{aligned}
\end{equation*}
To prove convergence here, it is useful to note that in addition to
\eqref{eq:neumann},  the series
\begin{equation*}
    \e^{\delta |x|} R_\phi(\lambda + \im \eta) \e^{-\delta |x|}
    \ = \ - \frac{1}{\lambda}  \sum_{n=0}^\infty
         \left [ \frac{\e^{\delta |x|} H_\phi \e^{-\delta |x|} -
         \im \eta}{\lambda} \right
        ]^n
\end{equation*}
is also absolutely convergent, in light of \eqref{eq:1}.

By cyclicity of the trace
\begin{equation*}
\begin{aligned}
    \tr T_\phi(\lambda + \im \eta) \ = \ -
    \sum_{N=0}^\infty \frac{1}{ \lambda^{N+3}} & \sum_{n =
    0}^N (n+1) \tr  \left ( H_\phi - \im \eta \right )^{n}
    \com{H_\phi}{\Lambda_1} \cdot \\ & \quad  \quad \quad
    \cdot \left ( H_\phi - \im \eta \right )^{N-n}
    \com{H_\phi}{\Lambda_2} \; ,
\end{aligned}
\end{equation*}
and, making use of the identity $\overline {\tr T} = \tr T^*$,
\begin{equation*}
\begin{aligned}
    \overline{\tr T_\phi(\lambda - \im \eta)} \ = \ -
    \sum_{N=0}^\infty \frac{1}{ \lambda^{N+3}} & \sum_{n =
    0}^N (N-n+1) \tr  \left ( H_\phi - \im \eta \right )^{n}
      \com{H_\phi}{\Lambda_1}  \cdot \\  & \quad  \quad \quad
    \cdot \left ( H_\phi - \im \eta \right )^{N-n}
     \com{H_\phi}{\Lambda_2} \; .
\end{aligned}
\end{equation*}
Thus
\begin{multline*}
     \tr T_\phi(\lambda + \im \eta)  - \overline{\tr T_\phi(\lambda - \im \eta)}
     \\ = \ -
    \sum_{N=0}^\infty \frac{1}{ \lambda^{N+3}}  \sum_{n =
    0}^N (2n -N) \tr  \left ( H_\phi - \im \eta \right )^{n}
      \com{H_\phi}{\Lambda_1}
    \cdot \left ( H_\phi - \im \eta \right )^{N-n}
     \com{H_\phi}{\Lambda_2} \; ,
\end{multline*}
which is the desired expansion.

The first term ($N=0$) of this series vanishes trivially. The second ($N=1$)
also vanishes, because
\begin{multline}\label{eq:N=1}
    \tr \com{H_\phi}{\Lambda_1} \left ( H_\phi - \im \eta \right )
     \com{H_\phi}{\Lambda_2} - \tr \left ( H_\phi - \im \eta \right ) \com{H_\phi}{\Lambda_1}
     \com{H_\phi}{\Lambda_2} \\
    \begin{aligned}
     &= \ - \tr
     \com{H_\phi}{\com{H_\phi}{\Lambda_1}} \com{H_\phi}{\Lambda_2}
     \\ &= \ - \sum_{x} \sum_y
     \com{H_\phi}{\com{H_\phi}{\Lambda_1}}(x,y)
     \com{H_\phi}{\Lambda_2}(y,x) \ = \ 0 \; ,
     \end{aligned}
\end{multline}
since $\com{H_\phi}{\Lambda_2}(y,x) \neq 0$ only for $|x-y|=1$ and
$\com{H_\phi}{\com{H_\phi}{\Lambda_1}}(x,y) \neq 0$ only for $|x-y|=0,2$ as
only nearest neighbor hopping terms are present in $H_\phi$. However the
coefficient of $\lambda^{-5}$ ($N=2$) is non-zero, and given by
\begin{multline*}
    2\tr \com{H_\phi}{\Lambda_1} \left ( H_\phi - \im \eta \right
    )^2
     \com{H_\phi}{\Lambda_2} - 2 \tr \left ( H_\phi - \im \eta \right )^2
     \com{H_\phi}{\Lambda_1}
     \com{H_\phi}{\Lambda_2} \\
    \begin{aligned}
     &= \  2\tr \com{H_\phi}{\Lambda_1}  H_\phi^2
     \com{H_\phi}{\Lambda_2} - 2 \tr H_\phi^2 \com{H_\phi}{\Lambda_1}
     \com{H_\phi}{\Lambda_2}\\ &= \ - 2 \tr H_\phi^2
     \com{\com{H_\phi}{\Lambda_1}}{\com{H_\phi}{\Lambda_2}} \; ,
     \end{aligned}
\end{multline*}
since the term proportional to $\eta$ vanishes by \eqref{eq:N=1} and the term
proportional to $\eta^2$ is the trace of a commutator, $
    \tr \com{\com{H_\phi}{\Lambda_1}}{\com{H_\phi}{\Lambda_2}} = 0
    $.

To calculate this term explicitly, recall that $\Lambda_i=I[x_i <0]$ so, by
\eqref{eq:Harper},
\begin{equation*}
\begin{aligned}
\com{H_\phi}{\Lambda_1}(x,x') \ &= \ (\Lambda_1(x') - \Lambda_1(x))
H_\phi(x,x')
\\ &= \
\begin{cases}
      1 \;, & x=(0,x_2) \; , \  x' = (-1,x_2) \; , \\
      -1 \; , & x=(-1,x_2) \; , \  x' = (0,x_2) \; , \\
      0 \; , & \text{all other $x,x'$,}
    \end{cases}
\end{aligned}
\end{equation*}
which is more succinctly expressed in Dirac notation:
\begin{equation*}
    \com{H_\phi}{\Lambda_1} \ = \ \sum_{a \in \Z}\ket{0,a}\bra{-1,a}-
    \ket{-1,a}\bra{0,a}\; .
\end{equation*}
Similarly,
\begin{equation*}
    \com{H_\phi}{\Lambda_2} \ = \ \sum_{a \in \Z} \e^{\im \phi a}
    \ket{a,0}\bra{a,-1}- \e^{-\im \phi
    a} \ket{a,-1}\bra{a,0}\; .
\end{equation*}
Thus
\begin{multline*}
    \com{\com{H_\phi}{\Lambda_1}}{\com{H_\phi}{\Lambda_2}} \ = \ (\e^{-\im
    \phi}-1) \Bigl ( \, \ket{0,0}\bra{-1,-1}  +  \ket{-1,0} \bra{0,-1} \, \Bigr  ) \\
     - (\e^{\im \phi}- 1 ) \Bigl ( \, \ket{0,-1}\bra{-1,0} +
     \ket{-1,-1}\bra{0,0} \,
     \Bigr ) \; ,
\end{multline*}
and
\begin{multline*}
    \tr H_\phi^2
    \com{\com{H_\phi}{\Lambda_1}}{\com{H_\phi}{\Lambda_2}} \\ = \
    (\e^{-\im \phi} -1 ) \Bigl ( \bra{-1,-1} H_\phi^2 \ket {0,0}
    + \bra{0,-1} H_\phi^2 \ket{-1,0}\Bigr ) \ - \ c. c. \; .
\end{multline*}
Finally, since
\begin{equation*}
  \bra{-1,-1} H_\phi^2 \ket {0,0} \ = \ 1 + \e^{-\im \phi} \, ,
  \quad \bra{0,-1} H_\phi^2 \ket{-1,0} \ = \  1 + \e^{\im \phi}
  \; ,
\end{equation*}
we have
\begin{equation*}
\begin{aligned}
   2 \tr H_\phi^2 \com{\com{H_\phi}{\Lambda_1}}{\com{H_\phi}{\Lambda_2}}
    \ &= \  4 (\e^{-\im
    \phi}-1) \left ( \cos(\phi) + 1  \right ) \ - \ c.c.
    \\ &= \ -8\im  \sin(\phi) ( \cos(\phi) + 1)  \; .
\end{aligned}
\end{equation*}
Therefore
\begin{gather*}
     \tr T_\phi(\lambda + \im \eta)  - \overline{\tr T_\phi(\lambda - \im \eta)} \ = \
      8\im  \sin(\phi) ( \cos(\phi) + 1) \lambda^{-5} \ + \
     \mathcal O(\lambda^{-6}) \; , \intertext{and}
     j_B(\lambda) \ = \  -\frac{4 \alpha}{\pi} \sin(\phi) ( \cos(\phi) + 1) \lambda^{-5}
     \ + \ \mathcal O(\lambda^{-6}) \; ,
\end{gather*}
which gives \eqref{eq:asymptotic}. This completes the proof of Theorem
\ref{thm:harper}. \qed
\appendix

\renewcommand{\theequation}{A.\arabic{equation}}
\renewcommand{\thesubsubsection}{A.\arabic{subsubsection}}
\section*{Appendix: conductance plateaus}\setcounter{equation}{0}
Localization is an essential prerequisite for the QHE.  Some localization
condition, valid at energies in an interval $\Delta$, is proven and used in
\cite{BvESB94,AG98}. It ensures that $\sigma_B(\lambda)$ is
\begin{enumerate}
  \item well defined as given by
  \eqref{eq:4},
  \item constant in $\lambda \in \Delta$, and
  \item $2 \pi \sigma_B(\lambda) \in \Z$.
\end{enumerate}
These results also rest on a homogeneity assumption for the Hamiltonian $H_B$,
or on its Fermi projections $P_\lambda$, namely that they be invariant or
ergodic under magnetic translations. The purpose of the Appendix is to
establish (1.-3.) under assumptions (\ref{eq:1}-\ref{eq:3}), which do not
entail translation invariance.

\begin{proposition}\label{prop:constant}
  Assume (\ref{eq:1}) and (\ref{eq:2}). Then $\sigma_B(\lambda)$ is
  well-defined. If in addition (\ref{eq:3}) holds, then $\sigma_B(\lambda)$ is
  constant in $\lambda \in \Delta$.
\end{proposition}

\begin{proposition}\label{prop:quantized}
  Assume (\ref{eq:1}) and (\ref{eq:2}). Then $2 \pi \sigma_B(\lambda) \in
  \Z$ for $\lambda \in \Delta$.
\end{proposition}

We remark that here constancy is proven without combining integrality and
continuity.

\subsubsection{Proof of Prop.\ \ref{prop:constant}} We consider Borel sets $S
\subset \R$ that either contain or are disjoint from $\set{\lambda | \lambda <
\Delta}$ and similarly for $\set{\lambda | \lambda > \Delta}$.  The class of
such sets $S$ is closed under unions and complements. We associate a bulk Hall
conductance to $S$ by setting
\begin{equation}\label{eq:A1}
\begin{aligned}
    \sigma_B(S) \ &=\ -\im \tr E_S
    \com{\com{E_S}{\Lambda_1}}{\com{E_S}{\Lambda_2}} \\
    &= \ \im \tr \left ( E_S \Lambda_1 E_S^\perp \Lambda_2 E_S -
    E_S \Lambda_2 E_S^\perp \Lambda_1 E_S \right ) \; ,
\end{aligned}
\end{equation}
where $E_S^\perp = 1 - E_S$ and the second line follows from $$ E_S
\com{E_S}{\Lambda_1} = E_S \com{E_S}{\Lambda_1} E_S^\perp = E_S \Lambda_1
E_S^\perp \; .$$ Note that $\sigma_B(\lambda_0) = \sigma_B((-\infty,
\lambda_0))$. We claim that, if $S_1 \cap S_2 = \emptyset $, then
\begin{gather} \label{eq:a}
     E_{S_1} \Lambda_1 E_{S_2} \Lambda_2 E_{S_1} \in
  \I_1 \; , \\ \label{eq:b}
    \sigma_B(S_1 \cup S_2) = \sigma_B(S_1) +
  \sigma_B(S_2) \; , \intertext{and moreover} \label{eq:c}
  \lim_{n \rightarrow \infty} \sigma_B(S_n) = 0 \text{ if } S_n
  \ \downarrow \ \emptyset \; .
\end{gather}
In particular, \eqref{eq:a} and its adjoint for $S_1 = S$, $S_2 = \R \setminus
S$ imply that the two terms in the final expression of \eqref{eq:A1} are
separately trace class.

\eqref{eq:a}: In the factorization
\begin{equation}
  \label{eq:A3}
  E_{S_1} \Lambda_1 E_{S_2} \Lambda_2 E_{S_1} \ = \ E_{S_1} \Lambda_1 E_{S_2}
  \e^{3 \delta |x_1|} \e^{-\delta |x|} \cdot \e^{-\delta |x|} \cdot \e^{-\delta
  |x|} \e^{3 \delta |x_2|} E_{S_2} \Lambda_2 E_{S_1} \; ,
\end{equation}
the middle $\e^{-\delta |x|} = \e^{-\delta |x_1|} \e^{-\delta |x_2|}$ is trace
class by \eqref{eq:44}, so that we need to show
\begin{equation}\label{eq:A4}
    \norm{E_{S_1} \Lambda_i E_{S_2} \e^{3 \delta |x_i|} \e^{- \delta |x|}} \ <
    \ \infty \;, \quad (i=1,2) \; .
\end{equation}
This follows from (\ref{eq:new3}, \ref{eq:new4}) and part (iii) of Lemma
\ref{lem:confinement}, with a bound which is uniform in $S_1$, $S_2$.

\eqref{eq:c}: By (\ref{eq:A1}, \ref{eq:A3}) and \eqref{eq:95} it suffices to
show
\begin{equation*}
    E_{S_n} \Lambda_i E_{S_n}^\perp \e^{3 \delta |x_i|} \e^{-\delta |x|} \
    \xrightarrow[n \rightarrow \infty]{s} \ 0 \;
    .
\end{equation*}
Since the l.h.s.\ is uniformly bounded in norm by the remark just made, we may
drop the exponentials as explained in connection with (\ref{eq:51},
\ref{eq:52}). Then the claim becomes obvious.

\eqref{eq:b}: From $E_{S_1 \cup S_2} = E_{S_1} + E_{S_2}$ and \eqref{eq:44a} we
have
\begin{equation*}
    \sigma_B(S_1 \cup S_2) \ = \ \sum_{i=1}^2 \left ( \tr E_{S_i}
    \Lambda_i E_{S_1 \cup S_2}^\perp \Lambda_2 E_{S_i} - \tr
    E_{S_1 \cup S_2}^\perp \Lambda_1 E_{S_i} \Lambda_2 E_{S_1 \cup
    S_2}^\perp \right ) \; .
\end{equation*}
We use $E_{S_1 \cup S_2}^\perp = E_{S_i}^\perp - E_{S_{i+1}}$ (with $i+1$
defined mod $2$) and obtain
\begin{equation*}
    \begin{aligned}
      \sigma_B(S_1 \cup S_2) \ & =  \ \sum_{i=1}^2 \sigma_B(S_i)
      - \sum_{i=1}^2 \tr E_{S_i} \Lambda_1 E_{S_{i+1}} \Lambda_2
      E_{S_i} \\
      & \qquad + \ \sum_{i=1}^2 \tr  E_{S_{i+1}}\Lambda_1 E_{S_i} \Lambda_2
      E_{S_{i+1}} \\
      &=  \ \sigma_B(S_1) + \sigma_B(S_2) \; .
    \end{aligned}
\end{equation*}

We finally prove constancy by showing that $\sigma_B([a,b]) = 0$ for any $[a,b]
\subset  \Delta$. Since $\sigma(H_B)$ is pure point in $\Delta$ we have
\begin{equation*}
    E_n \ := \ \sum_{i=1}^n E_{\set{\lambda_i}} \ \xrightarrow[n
    \rightarrow \infty]{s} \ E_{\mathcal E_{[a,b]}} \ = \
    E_{[a,b]} \; ,
\end{equation*}
where $\lambda_i$ is any labeling of the eigenvalues $\lambda \in \mathcal
E_{[a,b]}$. Now $E_n$ is a finite dimensional projection by \eqref{eq:3},
whence the two terms in
\begin{equation*}
    \sigma_B\left ( \cup_{i=1}^n \set{\lambda_i} \right ) \ = \ -
    \im \tr \left ( E_n \Lambda_1 E_n \Lambda_2 E_n - E_n
    \Lambda_2 E_n \Lambda_1 E_n \right ) \ = \ 0
\end{equation*}
are separately trace class.  They cancel by \eqref{eq:44a}. We conclude by
(\ref{eq:b}, \ref{eq:c}) that
\begin{equation*}
    \sigma_B([a,b]) \ = \ \sigma_B \left ( \cup_{i=1}^n \set{\lambda_i} \right )
    \ + \ \sigma_B\left (  \mathcal E_{[a,b]} \setminus
    \cup_{i=1}^n \set{\lambda_i} \right )
    \ \xrightarrow[n \rightarrow \infty]{} \ 0 \; . \quad \qed
\end{equation*}

\subsubsection{Proof of Prop.\ \ref{prop:quantized}} As
in \cite{ASS94} we are going to establish that $2 \pi \sigma_B(\lambda)$ is an
integer by relating it to the index of a pair of projections.

We first allow the functions $\Lambda_i$ in \eqref{eq:4} to switch values at
points other than the origin.  Let $p=(p_1,p_2) \in \Z^{2*} = \Z^2 + (\half,
\half)$ be the center of a plaquette and set
\begin{equation}\label{eq:A6}
\begin{split}
    \sigma_p \ &= \ - \im \tr P_\lambda
    \com{\com{P_\lambda}{\Lambda_{1,p}}}{\com{P_\lambda}{\Lambda_{2,p}}} \\ &= \
    \im \tr \left ( \com{P_\lambda}{\Lambda_{1,p}} P_\lambda^\perp
    \com{P_\lambda}{\Lambda_{2,p}} - \com{P_\lambda}{\Lambda_{2,p}} P_\lambda^\perp
    \com{P_\lambda}{\Lambda_{1,p}} \right ) \; ,
\end{split}
\end{equation}
where $\Lambda_{i,p} = \Lambda(x_i - p_i)$, $(i=1,2)$. (Since $\Lambda(n) =
\Lambda(n+\half)$ for $n \in \Z$, $\sigma_B(\lambda)$ is just $\sigma_p$ for
$p=-(\half,\half)$.)

To define the index, let $\theta_p(x) = \arg(x-p)$ be the angle of sight of $x
\in \Z^2$ from $p$, and set $U_p(x) = \e^{\im \theta_p(x)}$.  The relevant
index is $N_p = \Ind(U_p P_\lambda U_p^*, P_\lambda)$, where $\Ind(P,Q)$
denotes the index of a pair of projections introduced in ref.~\cite{ASS94}:
\begin{equation}\label{eq:Index}
    \Ind(P,Q) \ := \ \dim \ran P \cap \ker Q - \dim \ran Q \cap \ker P \; .
\end{equation}
We recall the following basic properties of $\Ind(\cdot,\cdot)$:
\begin{enumerate}
  \item If $P-Q$ is compact, $\Ind(P,Q)$ is well defined and finite.
  \item If $(P-Q)^{2n+1}$ is trace class for some integer $n \ge 0$, then
  \begin{equation}\label{eq:IndexasTrace}
    \tr (P-Q)^{2n+1} = \Ind(P,Q) \; .
  \end{equation}
\end{enumerate}

Since $N_p$ is an integer by \eqref{eq:Index}, Prop.~\ref{prop:quantized} is a
consequence of the identity
\begin{equation*}
    2 \pi \sigma_B(\lambda) =  N_p \; ,
\end{equation*}
to be proved below. Indeed, this is the same strategy employed in
refs.~\cite{ASS94,AG98}. The starting point for our proof is the observation
that $\sigma_p$ and $N_p$ are independent of $p$ even without ergodicity for
the underlying projection.
\begin{lemma}\label{lem:translationinvariance}
  The index $N_p$ is well defined for any $p \in \Z^{2*}$, and for any $a \in
  \Z^2$
  \begin{enumerate}
    \item[i)] $N_{p+a} = N_p$,
    \item[ii)] $\sigma_{p+a} = \sigma_p$.
  \end{enumerate}
\end{lemma}
\begin{proof}
  Part (i) follows from \cite[Prop. 3.8]{ASS94} once we verify that $N_p$ is
  well defined.  For this we follow \cite{AG98} and show that $(P_\lambda - U_p
  P_\lambda U_p^*)^3$ is trace class, using
  \begin{lemma*}[{\cite[Lemma 1]{AG98}}] For an operator with the matrix
  elements $T_{x,y}$
    \begin{equation*}
    \norm{T}_3 \ \equiv \ (\tr |T|^3)^{1/3} \ \le \ \sum_{b}\left ( \sum_x
    |T_{x+b,x}|^3 \right )^{1/3} \; .
    \end{equation*}
  \end{lemma*}
  In our case, with $T = P_\lambda - U_p
  P_\lambda U_p^*$, we have (see \cite[eq. (4.13)]{AG98})
  \begin{multline*}
    |T(x+b,x)| \ = \ |1- \e^{\im (\theta_p(x+b) - \theta_p(x))}|
    |P_\lambda(x+b,x)| \\
    \le \ C \frac{|b|}{1+ |x-p|} |P_\lambda(x+b,x)|  \ \le \
     C(1+|p|) \frac{|b|}{1 + |x|} |P_\lambda(x+b,x)| \; .
  \end{multline*}
  (Here and in the sequel,
  $C$ denotes a generic constant, whose value
  is independent of any lattice sites in the given inequality, though
  that value may change from line
  to line.)

  Since \eqref{eq:2} holds for $g(H_B) = P_\lambda$, we have
  \begin{equation*}
    |P_\lambda(x+b,x)| \ \le \ C_2 (1+ |x|)^\nu \e^{-\mu |b|} \; ,
  \end{equation*}
  but we also have $|P_\lambda(x+b,x)| \le 1$, because $\norm{P_\lambda} \le 1$.
  Combing these two estimates gives
  \begin{equation}\label{eq:2optimized}
    |P_\lambda(x+b,x)| \ \le \ \begin{cases}
      1 & |b| \le \frac{2 \nu}{\mu} \ln(|x|+1)  \; ,\\
      C_2 \,  \e^{-\frac{\mu}{2} |b|}
      & |b| > \frac{2 \nu}{\mu} \ln(|x|+1) \; .
    \end{cases}
  \end{equation}
  Thus
  \begin{multline*}
    \left ( \sum_{x} |T(x+b,x)|^3 \right )^{1/3} \\
    \le \ C (1+|p|) |b| \left (
        \sum_{|x| < \e^{\frac{\mu}{2 \nu} |b|} - 1} \frac{\left[ C_2
        \e^{-\frac{\mu}{2} |b|} \right ]^3}{(1 + |x|)^3} \ + \
        \sum_{|x| \ge \e^{\frac{\mu}{2 \nu} |b|} - 1} \frac{1}{(1 + |x|)^3}
        \right )^{1/3} \\
     \le \ C (1+|p|) |b| \left ( \e^{-\frac{\mu}{2} |b|} \ + \
    \e^{-\frac{\mu}{6 \nu} |b|} \right ) \; .
  \end{multline*}
  Since the last line is clearly summable over $b$, we see that $(U_p P_\lambda
  U_p^* - P)^3$ is trace class, and therefore the index $N_p$ is well defined.

  Turning now to part (ii), we note that we may just treat the case
  $p=-(\half,\half)$, $a=(a_1,0)$, the case of translation in the 2-direction
  being similar.  By (\ref{eq:A1}, \ref{eq:a}, \ref{eq:44a}) we need to show that
  \begin{multline}\label{eq:A7}
    \tr (P_\lambda (\Delta \Lambda_1) P_\lambda^\perp \Lambda_2 P_\lambda)
    - \tr (P_\lambda \Lambda_2 P_\lambda^\perp (\Delta \Lambda_1) P_\lambda )
    \\
    = \ \tr (P_\lambda (\Delta \Lambda_1) P_\lambda^\perp \Lambda_2 P_\lambda)
    - \tr (P_\lambda^\perp (\Delta \Lambda_1) P_\lambda  \Lambda_2 P_\lambda^\perp )
\end{multline}
vanishes, where $\Delta \Lambda_1(x) = \Lambda(x_1) - \Lambda(x_1 - a_1)$ is
compactly supported in $x_1$. We claim that $(\Delta\Lambda_1) P_\lambda^\perp
\Lambda_2 P_\lambda \in \I_1$. This follows like \eqref{eq:a}  through the
factorization
\begin{equation*}
    (\Delta\Lambda_1) P_\lambda^\perp \Lambda_2 P_\lambda
    \ = \ (\Delta\Lambda_1)\e^{3 \delta |x_1|} \e^{-\delta |x|} \cdot \e^{-\delta |x|}
    \cdot \e^{-\delta |x|} \e^{3 \delta |x_2|} P_\lambda^\perp \Lambda_2
    P_\lambda \; ,
\end{equation*}
by noticing that the first factor, which is new, is bounded.  Likewise
$$(\Delta \Lambda_1) P_\lambda \Lambda_2 P_\lambda^\perp \in \I_1 \; .$$
Therefore \eqref{eq:A7} equals
\begin{equation*}
    \tr  (\Delta\Lambda_1) P_\lambda^\perp \Lambda_2 P_\lambda
    - \tr (\Delta \Lambda_1) P_\lambda \Lambda_2 P_\lambda^\perp \ = \ \tr
    (\Delta \Lambda_1) \com{\Lambda_2}{P_\lambda} \ = \ 0 \; ,
\end{equation*}
by evaluating the trace in the position basis. \qed
\end{proof}

The proof of Prop.~\ref{prop:quantized} is now completed by the following
result, with the translation invariance required in the argument of
\cite{ASS94} now provided by Lemma~\ref{lem:translationinvariance}.
\begin{lemma}
  Let $\Lambda_L = \set{-L,\ldots, L}^2 \subset \Z^2$.  Then
  \begin{equation}\
    \label{eq:A8}
    \left . \begin{array}{c}
    N/ 2 \pi \\
    \sigma_B(\lambda) \end{array} \right \} \ = \
    \lim_{L \rightarrow \infty} \frac{-2 \im}{(2L + 1)^2} \sum_{\substack{y,z
    \in \Z^2 \\ x \in \Lambda_L}} P_\lambda(x,y) P_\lambda(y,z) P_\lambda(z,x)
    \Area(x,y,z) \; ,
  \end{equation}
  where $N$, resp.~$\sigma_B(\lambda)$ are the translation invariant values of
  $N_p$, resp.~$\sigma_p$, and $\Area(x,y,z)$ is the triangle's oriented area,
  namely $\half (x-y)\wedge (y-z)$.
\end{lemma}
\begin{remark}
  The r.h.s.\ of \eqref{eq:A8} is the trace per unit
  volume of
  \begin{equation*}
   -\im P_\lambda \com{\com{P_\lambda}{X_1}}{\com{P_\lambda}{X_2}} \; ,
   \end{equation*}
   which may be interpreted as the macroscopic version of \eqref{eq:4}.
\end{remark}

\begin{proof}
  The first statement makes use of Connes' area formula \cite{Co86} in the
  version \cite{ASS94} adapted to the lattice \cite{AG98}:
  \begin{quotation}
    For a fixed triplet $u^{(1)}$, $u^{(2)}$, $u^{(3)} \in \Z^2$, let
    $\alpha_i(p) \in (-\pi, \pi)$ be the angle of view from $p \in \Z^{2*}$ of
    $u^{(i+2)}$ relative to $u^{(i+1)}$ (with $\alpha_i(p) =0$ if $p$ lies
    between them). Then
    \begin{equation}\label{eq:A9}
    \sum_{p \in \Z^{2*}}\sum_{i=1}^3 \sin \alpha_i(p) \ = \ 2 \pi
    \Area(u^{(1)}, u^{(2)}, u^{(3)}) \; .
    \end{equation}
  \end{quotation}
  By the computation of \cite{ASS94},
  \begin{equation*}
    N_p \ = \ \tr (U_p P_\lambda U_p - P_\lambda )^3 \ = \ - 2 \im \sum_{x,y,z
    \in \Z^2} P_\lambda(x,y) P_\lambda(y,z) P_\lambda(z,x)  S(p,x,y,z)  \; .
\end{equation*}
with $S(p,x,y,z) = \sin\angle(x,p,y) + \sin\angle(y,p,z) + \sin\angle(z,p,x) $.
Letting $\Lambda_L^* = \set{-L+\half, \ldots, L + \half}^2 \subset \Z^{2*}$ we
have that $N (2L +1)^2$ is the sum of the r.h.s.\ over $p \in \Lambda_L^*$.

We would like to replace the sum over $x \in \Z^2$, $p \in \Lambda_L^*$ by that
over $x \in \Lambda_L$, $p \in \Z^{2*}$. The error is estimated by
\begin{equation}\label{eq:errorterm}
    \sum_{ \substack{x \in \Z^2 \setminus \Lambda_L \\
        p \in \Lambda_L^*}}
        \abs{f(p,x)}
    + \sum_{\substack{x \in \Lambda_L \\
    p \in \Z^{2*} \setminus \Lambda_L^*}}
        \abs{f(p,x)} \; ,
\end{equation}
where
\begin{equation*}
    f(p,x) \ := \ -2 \im \sum_{y,z
    \in \Z^2} P_\lambda(x,y) P_\lambda(y,z) P_\lambda(z,x) S(p,x,y,z) \; .
\end{equation*}

By \eqref{eq:2} for $g(H_B) =P_\lambda$ the points $y,z$ are exponentially
clustered around $x$, so we have $|f(p,x)| \le C_x (1 + |p-x|)^{-3}$.  However
because of the pre-factor $(1 + |x|)^\nu$ in \eqref{eq:2}, the constant $C_x$
carries some dependence on $x$ (as indicated), which must be controlled in
order to bound \eqref{eq:errorterm}.

In fact, the following estimate for $|f(p,x)|$ is true:
\begin{equation}\label{eq:fpxbound}
  \abs{f(p,x)} \ \le \ C \frac{[1 + \ln (1 + |x|)]^5}{1+|x-p|^3} \; .
\end{equation}
Before proving \eqref{eq:fpxbound}, let us see how it allows us to complete the
proof.  Indeed, since
\begin{equation*}
    \sum_{\substack{x \in \Lambda_L \\
    p \in \Z^{2*} \setminus \Lambda_L^*}} \frac{1}{(1 + |x-p|)^3} \ = \
    \mathcal O \left ( L \ln L \right ) \; , \quad L \rightarrow \infty \; ,
\end{equation*}
as far as the second term of \eqref{eq:errorterm} is concerned, we have
\begin{equation*}
    \sum_{\substack{x \in \Lambda_L \\
    p \in \Z^{2*} \setminus \Lambda_L^*}}
        \abs{f(p,x)} \ \le \ C [\ln L]^5  \sum_{\substack{x \in \Lambda_L \\
    p \in \Z^{2*} \setminus \Lambda_L^*}} \frac{1}{(1 + |x-p|)^3} \ = \
    \mathcal{O}( L [\ln L]^6 )
        \;.
\end{equation*}
For the first term we note that
\begin{equation*}
    [1 + \ln (1 + |x|)]^5 \ \le \ C  (\ln L )^5 [1 + \ln (1 + |x-p|)]^5 \; ,
\end{equation*}
for $x, p$ in the indicated range and large $L$, resulting in
\begin{equation*}
    \sum_{ \substack{x \in \Z^2 \setminus \Lambda_L \\
        p \in \Lambda_L^*}}
        \abs{f(p,x)} \ \le \ C [\ln L]^5  \sum_{\substack{p \in \Lambda_L^* \\
     x\in \Z^{2} \setminus \Lambda_L}} \frac{[1 + \ln (1 + |x-p|)]^5 }{(1 + |x-p|)^3} \ = \
    \mathcal{O}( L [\ln L]^{11} ) \; .
\end{equation*}
Therefore,
\begin{multline*}
    N (2L+1)^2 \ = \ \sum_{\substack{x \in \Lambda_L \\ p \in \Z^{2*}}} f(p,x)
     \ + \ \mathcal O(L [\ln L]^{11}) \\
     = \ - 2 \im \sum_{\substack{x \in \Lambda_L \\ y,z
    \in \Z^2}} P_\lambda(x,y) P_\lambda(y,z) P_\lambda(z,x)
    \sum_{p \in \Z^{2*}}S(p,x,y,z) \ + \ \mathcal O(L [\ln L]^{11}) \; ,
\end{multline*}
which gives \eqref{eq:A8} for $N/2 \pi$ after applying Connes' area formula and
taking the limit $L \rightarrow \infty$.

As for the proof of \eqref{eq:fpxbound}, we consider separately the cases (i)
$|p-x| < \frac{2\nu}{\mu} \ln(|x| +1)$ and (ii) $|p-x| \ge \frac{2\nu}{\mu}
\ln(|x| +1)$. In case (i), we use the bound $|S(p,x,y,z)| \le 3$ to conclude
\begin{equation*}
  |f(p,x)| \ \le \ 6 \sum_{y,z
    \in \Z^2} \abs{P_\lambda(x,y) P_\lambda(y,z) P_\lambda(z,x)} \
    \le \ 6 \sum_{y \in \Z^2} \abs{P_\lambda(x,y)} \; ,
\end{equation*}
since
\begin{multline*}
    \sum_{z \in \Z^2} \abs{P_\lambda(y,z) P_\lambda(z,x)}
    \ \le \ \left ( \sum_{z \in \Z^2} \abs{P_\lambda(y,z)}^2 \sum_{z \in \Z^2}
    \abs{P_\lambda(z,x)}^2 \right )^{1/2} \\
    \le \ \left [ P_\lambda(y,y) P_\lambda(x,x) \right ]^{1/2} \ \le \ 1 \; .
\end{multline*}
Now by \eqref{eq:2optimized},
\begin{multline*}
    \sum_{y \in \Z^2} \abs{P_\lambda(x,y)} \ \le \ \left (4 \frac{\nu}\mu \ln(|x| + 1) +
    1 \right )^2 \ + \ C_2 \sum_{|b| > \frac{2 \nu}{\mu} \ln (|x| + 1)}
    \e^{-\frac{\mu}{2} |b|} \\
    \le \ C \left [ 1 + \ln(|x| + 1) \right ]^2 \ \le \ C
    \frac{\left [ 1 + \ln(|x| + 1) \right ]^5}{(1 + |x - p|)^3} \; ,
\end{multline*}
where in the last step we have used that $|x-p| \le \frac{2\nu}{\mu} \ln(|x| +
1)$. This implies \eqref{eq:fpxbound} in case (i). To prove \eqref{eq:fpxbound}
in case (ii), consider separately the contributions to $f(p,x)$ coming when
both $y$ and $z$ fall inside the ball of radius $|p-x|$ around $x$ and when one
of $y$ or $z$ falls outside the ball. The latter contribution is exponentially
small in $|x-p|$, since it is bounded by
\begin{multline*}
    6 \left [ \sum_{\substack{|y-x| \ge |p-x| \\ z \in \Z^2}} +
    \sum_{\substack{|z-x| \ge |p-x| \\ y \in \Z^2}}
    \abs{P_\lambda(x,y) P_\lambda(y,z) P_\lambda(z,x)} \right ] \\
    \ \le \ 12 \sum_{|y-x| \ge |p-x|} \abs{P_\lambda(x,y) } \ \le \ C
    \e^{-\frac{\mu}{2} |x-p|} \; ,
\end{multline*}
where in the last step we have used \eqref{eq:2optimized} and the fact that
$|x-p| > \frac{2\nu}{\mu} \ln(|x| +1)$. To bound the former contribution note
that in this case both $\abs{\angle (y,p,x)}$ and $\abs{\angle (z,p,x)}$ are
smaller than $\frac{\pi}{2}$, and make use of the following estimates: (1)
given $\alpha, \beta \in (-\frac{\pi}{2},\frac{\pi}{2})$,
\begin{equation*}
    \abs{\sin \alpha + \sin \beta - \sin(\alpha + \beta)} \ \le \
    \abs{\sin\alpha}^3 + \abs{\sin \beta}^3 \; ,
\end{equation*}
and (2) given $y$ with $|y-x| < |p-x|$,
\begin{equation*}
    \abs{\sin \angle (y,p,x)} \ \le \  \frac{|y-x|}{1+|p-x|} \; .
\end{equation*}
Putting these two estimates together gives the following bound for the
contribution with $y,z$ in the ball of radius $|x-p|$ around $x$
\begin{multline*}
     \frac{C}{(1+|p-x|)^3} \sum_{|y-x|,|z-x| < |p-x|}
    \abs{P_\lambda(x,y) P_\lambda(y,z) P_\lambda(z,x)}
    \left ( |y-x|^3 + |z-x|^3 \right ) \\
    \le   \frac{C}{(1+|p-x|)^3} \sum_{y \in \Z^2}
    \abs{P_\lambda(x,y)} |y-x|^3  \ \le \ C \frac{[1 + \ln(|x| + 1)]^5}
        {(1+|p-x|)^3} \; ,
\end{multline*}
where in the last step we have used \eqref{eq:2optimized}. This proves
\eqref{eq:fpxbound} in case (ii) and completes the proof of \eqref{eq:A8} for
$N/2 \pi$.

The proof for $\sigma_B$ is similar.  By evaluating \eqref{eq:A6} in the
position basis as in \cite{ASS94} we obtain
\begin{multline}\label{eq:A10}
    \sigma_p \ = \ \im \sum_{x,y,z \in \Z^2} P_\lambda(x,y) P_\lambda^\perp(y,z)
    P_\lambda(z,x) \cdot \\
    \cdot\left [ (\Lambda(y_1 - p_1) - \Lambda(x_1-p_1) )
    (\Lambda(z_2 - p_2) - \Lambda(y_2 - p_2) ) - (1 \leftrightarrow 2)\right ]
    \; .
\end{multline}
We then sum over $p \in \Lambda_L^*$ and move the anchor from $p$ to $x$ (in
this case the corresponding $f(p,x)$ decays exponentially in $|p-x|$, again
with logarithmic growth in $|x|$). The sum over $p \in \Z^{2*}$ of the square
bracket in \eqref{eq:A10} involves
\begin{equation*}
    \sum_{p_i \in \Z^*} (\Lambda(y_i - p_i) - \Lambda (x_i - p_i)) \ = \ x_i -
    y_i
\end{equation*}
and thus equals $(x_1 -y_1)(y_2 - z_2) - (x_2 - y_2) (y_1 - z_1) = 2
\Area(x,y,z)$.  The proof is completed by $P_\lambda^\perp(y,z) = \delta_{yz} -
P_\lambda(y,z)$. \qed
\end{proof}


\end{document}